\documentclass{article}

\usepackage{arxiv}

\usepackage[utf8]{inputenc} 
\usepackage[T1]{fontenc}    
\usepackage{hyperref}       
\usepackage{url}            
\usepackage{booktabs}       
\usepackage{amsfonts}       
\usepackage{nicefrac}       
\usepackage{microtype}      
\usepackage{lipsum}
\usepackage{graphicx}
\graphicspath{ {./images/} }
\usepackage{epstopdf, epsfig}
\usepackage{caption}
\usepackage{color, soul}
\usepackage[colorinlistoftodos]{todonotes}

\usepackage{longtable} 
\usepackage{multirow} 
\usepackage{amsmath} 
\usepackage[T1]{fontenc} 
\usepackage{tabularx} 
\usepackage{changepage} 
\usepackage{natbib} 
\usepackage{float} 
\usepackage[running]{lineno} 
\usepackage{xcolor} 
\usepackage{relsize} 
\usepackage{url} 
\usepackage{appendix} 
\title{\rm{Persistence behaviour of heat and momentum fluxes in convective surface layer turbulence}}

\author{
Subharthi Chowdhuri \\
  Indian Institute of Tropical Meteorology\\
  Ministry of Earth Sciences\\
  Dr. Homi Bhaba Road, Pashan, Pune-411008, India\\
  \texttt{subharthi.cat@tropmet.res.in} \\
  \And
  Thara V Prabha \\
  Indian Institute of Tropical Meteorology\\
  Ministry of Earth Sciences\\
  Dr. Homi Bhaba Road, Pashan, Pune-411008, India\\
  \texttt{thara@tropmet.res.in} \\
  \And
Tirtha Banerjee \\
Department of Civil and Environmental Engineering\\
University of California, Irvine, CA 92697, USA \\ 
 \texttt{tirthab@uci.edu} \\
}

\begin{document}
\maketitle
\begin{abstract}
The characterization of heat and momentum fluxes in wall-bounded turbulence is of paramount importance for a plethora of applications, ranging from engineering to Earth sciences. However, how the turbulent structures associated with velocity and temperature fluctuations interact to produce the emergent flux signatures, is not evident till date. In this work, we investigate this fundamental issue by studying the switching patterns of intermittently occurring turbulent fluctuations from one state to another, a phenomenon called persistence. We discover that the persistence patterns for heat and momentum fluxes are widely different. Moreover, we uncover power-law scaling and length scales of turbulent motions that cause this behavior. Furthermore, by separating the phases and amplitudes of flux events, we explain the origin and differences between heat and momentum transfer efficiencies in convective turbulence. Our findings provide new understanding on the connection between flow organization and flux generation mechanisms, two cornerstones of turbulence research.
\end{abstract}


\section{Introduction}
\label{Intro}
The ensemble averaged vertical turbulent fluxes of momentum or heat are expressed as the covariance between the vertical and streamwise velocity fluctuations ($\overline{u^{\prime}w^{\prime}}$) or between the vertical velocity and temperature fluctuations ($\overline{w^{\prime}T^{\prime}}$). The primes in the flux expressions denote the turbulent fluctuations in the streamwise velocity ($u^{\prime}$), vertical velocity ($w^{\prime}$), or temperature ($T^{\prime}$). Besides, the overline symbol indicates the average over several ensemble members. For practical purposes, such ensemble average is replaced by the average over time or space by applying the ergodic hypothesis \citep{monin2013statistical}. In turbulent flows, these fluxes quantify the amount of heat or momentum being transported from (to) the surface to (from) other locations. The sign convention dictates that, the positive or negative values of the fluxes denote the direction of the transport from or towards the surface. The estimation of these fluxes have numerous usages, such as, in wall-bounded turbulent flows, the momentum transport towards the wall is related to surface drag which determines the power requirements and efficiencies in many engineering applications \citep[e.g.,][]{holmes2012turbulence}. Additionally, in the geophysical context, these fluxes quantify the surface-atmosphere momentum and heat exchanges, which eventually drive the Earth's climate \citep[e.g.,][]{baldocchi2001fluxnet}. Therefore, it is of paramount importance to develop a comprehensive understanding of the turbulent generation mechanisms of the heat and momentum fluxes. Since the flux computation involves the product of two turbulent quantities, a fundamental research question is, \emph{how do the turbulent structures (of different time or length scales) associated with the velocity and temperature fluctuations interact to produce the momentum and heat flux signatures}? 

The common method of describing the fluctuation characteristics and the associated fluxes, corresponding to the different scales of the turbulent motions, is the spectral approach \citep{Tay38}. However, a landmark study by \citet{kline1967structure} noted that the transport of momentum in wall-bounded shear flows was not a continuous process, as assumed by the spectral analysis. Instead they found that, the intermittent ejections of low-momentum fluid parcels from the wall ($u^{\prime}<0$ and $w^{\prime}>0$) accompanied with the sweeps of high-momentum fluid towards the wall ($u^{\prime}>0$ and $w^{\prime}<0$) were responsible for the generation of the momentum flux. Subsequently, to diagnose the intermittent signatures of the ejection and sweep motions and estimate their contributions towards the momentum flux, a conditional sampling technique named quadrant analysis was introduced \citep{wallace1972wall,lu1973measurements,wallace2016quadrant}. Along with the utility of the quadrant analysis, a few studies additionally noted that the ejection and sweep motions in wall-bounded turbulent flows occurred with a range of different time scales as more often than not they switched irregularly from one quadrant state to the other \citep{rao1971bursting,alfredsson1984time,luchik1987timescale}. While evaluating the statistical characteristics of these time scales, \citet{rao1971bursting} and \citet{alfredsson1984time} found that the mean time scales of the burst events (sequence of ejections exceeding a certain threshold) in the near-wall region were influenced by both the small and large scale motions. 

Such results were intriguing, considering their relevance to the connection between the turbulent structures of different time scales and the intermittent generation of the momentum fluxes. Nevertheless, a detailed treatment of the time scale distribution of the flux-carrying motions was severely lacking \citep{narasimha1990utility}. To tackle this problem, \citet{kailasnath1993zero} demonstrated that the probability density functions (PDFs) of the time scales of the turbulent motions contributing to the momentum flux can be systematically studied by exploring the zero-crossing properties of the instantaneous $u^{\prime}w^{\prime}$ signal. Their method was inspired by the results of \citet{sreenivasan1983zero} where the PDFs of the time intervals between the zero-crossings of the turbulent fluctuations were used to probe the distribution of time scales in a wall-bounded flow. Note that, in the parlance of non-equilibrium statistical mechanics, the distribution of zero-crossing time intervals in a stochastic signal is equivalent to the persistence PDFs, where persistence is the probability $P(t)$ that the signal does not change its sign up to the time $t$ \citep{majumdar1999persistence,chowdhuri2020persistence}. Hereafter, we refer the zero-crossing PDFs as the persistence PDFs, given its technical suitability \citep{chowdhuri2020persistence}.  

One of the crucial aspects of the study by \citet{kailasnath1993zero} was to investigate how the persistence PDFs of the individual velocity signals compared to the persistence PDFs of their product which constituted the momentum flux. They found that in the inertial layer of a flat-plate boundary layer, the persistence PDFs of the momentum flux signal closely followed the same PDFs of the vertical velocity fluctuations, with both displaying a nearly identical single exponential function. This was in sharp contrast with the streamwise velocity fluctuations, whose persistence PDFs displayed a double exponential structure with two different exponents. They interpreted this behaviour as both the small and large scales were relevant for the variations in the streamwise velocity fluctuations. However, for the variations in the wall-normal velocity and in the momentum flux, only the large scales were primarily responsible, since their zero-crossing PDFs displayed a single exponential function. Moreover, they also proposed a connection to associate such behaviour with the properties of the attached eddies which populated the inertial layer, consistent with the Townsend's attached eddy hypothesis \citep{townsend1980structure}. 

In general, the findings of \citet{kailasnath1993zero} shed light on the fundamental issue of the relationship between the turbulent structures of different scales and the associated intermittent flux signatures. However, while evaluating the time scales of the momentum flux signals, they did not consider the quadrant effects. We hypothesize that the conditional sampling of the flux events can establish an important link between the momentum flux partition among the four different quadrants and the time scales of the turbulent flow. Such an analysis can identify the time scales associated with individual quadrant motions that contribute towards momentum flux generation. To the best of our knowledge, there has been no subsequent studies to scrutinize this aspect. Although until now, we have discussed the characteristics of the momentum flux signal in turbulent shear flows, it can be noted that even in buoyancy-driven turbulent flows, the instantaneous heat flux signals ($w^{\prime}T^{\prime}$) display similar intermittent signatures \citep{shang2003measured}. Therefore, one can contemplate that, such investigation will be pertinent for a convective atmospheric surface layer (ASL) flow, where both the shear and buoyancy play significant roles to maintain the turbulence.

The ASL is a generalization of the inertial layer of wall-bounded shear flows by including the effect of buoyancy \citep{davidson2015turbulence}. The vertical extent of the ASL is approximately up to the lowest 10\% of the convective boundary layer (excluding the roughness sublayer) where the fluxes are nearly constant with height \citep{wyngaard1992atmospheric}. During the early days of ASL research, \citet{haugen1971experimental} observed that in convective conditions the time traces of the instantaneous $w^{\prime}T^{\prime}$ signals displayed intermittently occurring long sequences of positive heat flux events. They further noted that the time-averaged characteristics of the heat flux and their transport efficiencies (expressed through the correlation coefficient between $w^{\prime}$ and $T^{\prime}$) were primarily determined by these long persistent events. On the other hand, for the instantaneous $u^{\prime}w^{\prime}$ signals they found that the long sequences of the flux-carrying motions either transported momentum in the upward or in the downward direction. In a highly-convective ASL flow (characterized by strong thermal stratification), the amplitudes of the large persistent positive and negative momentum flux events were almost similar to each other. This caused the time-averaged momentum flux to become quite small due to the near-perfect cancellation of such positive and negative values. However, with the decrease in the strength of the thermal stratification (near-neutral ASL), the negative amplitudes associated with the long sequences overwhelmed the positive ones. Therefore, the momentum transport efficiency increased as the ASL approached the near-neutral conditions. \citet{haugen1971experimental} postulated that such changes associated with thermal stratification were related to the changes in the topology of coherent structures from cellular patterns to horizontal rolls. Nevertheless, their conclusions were based on visual inspections of a few 15-min records, and of a qualitative nature. 

Since then, numerous studies have documented that akin to laboratory flows, the heat and momentum transport processes in a convective ASL are intermittent in nature, which occur with a range of different time scales \citep{narasimha1990turbulent,duncan1992method,caramori1994structural,katul1994conditional,narasimha2007turbulent}. Additionally, several other researchers have demonstrated that the time-averaged transport characteristics of the heat and momentum in a convective ASL flow are strongly dependent on the strength of the thermal stratification \citep{de1993verification,li2011coherent,chowdhuri2019evaluation}. Specifically, these studies show that in a highly-convective ASL the heat and momentum transports are decoupled from each other, whereas they are strongly coupled in near-neutral conditions. Recently, \citet{salesky2017nature} have illustrated that, with thermal stratification (stability), the change in the topology of the coherent structures in a convective flow influences the transport efficiencies of the heat and momentum. 

Broadly, the characteristics of the time-averaged heat and momentum fluxes in a convective ASL flow could be perceived as a well-researched problem. However, since the times of \citet{haugen1971experimental}, a grand challenge still remains to connect the intermittent behaviour of these fluxes with the different scales of the turbulent eddies which influence their component signals. Undoubtedly, such investigation is of fundamental interest, to improve our understanding of the generation mechanisms of the heat and momentum fluxes. This problem also has practical implications regarding the development of next-generation transport models of convective turbulence. To address this issue, we ask the following questions:
\begin{enumerate}
    \item What type of event signatures are generated in the momentum and heat flux signals, due to the presence of turbulent eddies of different time scales?
    \item What is the relation between these event signatures at different time scales with the heat and momentum transport efficiencies in a convective ASL flow? 
    \item How does the change in the thermal stratification modulate the event signatures at different time scales?
\end{enumerate}
In this article, we attempt to answer these questions through persistence analysis of the turbulent heat and momentum fluxes in a convective ASL flow. The persistence analysis is well-suited for this purpose, since it deals with the time intervals between the zero-crossings of an intermittent turbulent signal, eventually being related to the time scales of the turbulent structures. During our presentation, we arrange the paper in three different sections. In Section \ref{Data}, we provide brief descriptions of the field-experimental dataset and methodology, in Section \ref{results} we introduce the results and discuss them, and lastly in Section \ref{conclusion} we summarize the key takeaways and provide the scope for further research.

\section{Data and Methodology}
\label{Data}
In this study, we have used the dataset from the Surface Layer Turbulence and Environmental Science Test (SLTEST) experiment. The details about the SLTEST experiment and the set-up of the instruments are provided in \citet{chowdhuri2019revisiting} and \citet{chowdhuri2020persistence}. In this experiment, nine North-facing time synchronized CSAT3 sonic anemometers were deployed on a 30-m mast, spaced logarithmically from 1.42 m to 25.7 m, with the sampling frequency being set at 20 Hz. We followed the same data processing steps as outlined in \citet{chowdhuri2020persistence} to select the 30-min periods from the daytime convective conditions. Total 261 combinations of the stability ratios ($-\zeta=z/L$, where $z$ is the measurement height and $L$ is the Obukhov length) were possible for these selected 30-min periods. The entire range of $-\zeta$ (12 $\leq -\zeta \leq$ 0.07) was divided into six stability classes and considered for the persistence analysis of the turbulent heat and momentum fluxes. These are the same set of stability classes used by \citet{chowdhuri2019revisiting} and \citet{chowdhuri2020persistence} for their study of turbulence anisotropy and persistence behaviour of turbulent velocity and temperature fluctuations (see their respective tables 2 and 1). Note that, for the sake of brevity, the same information is not repeated here. 

\begin{figure*}[h]
\centering
\hspace*{-1.2in}
\includegraphics[width=1.3\textwidth]{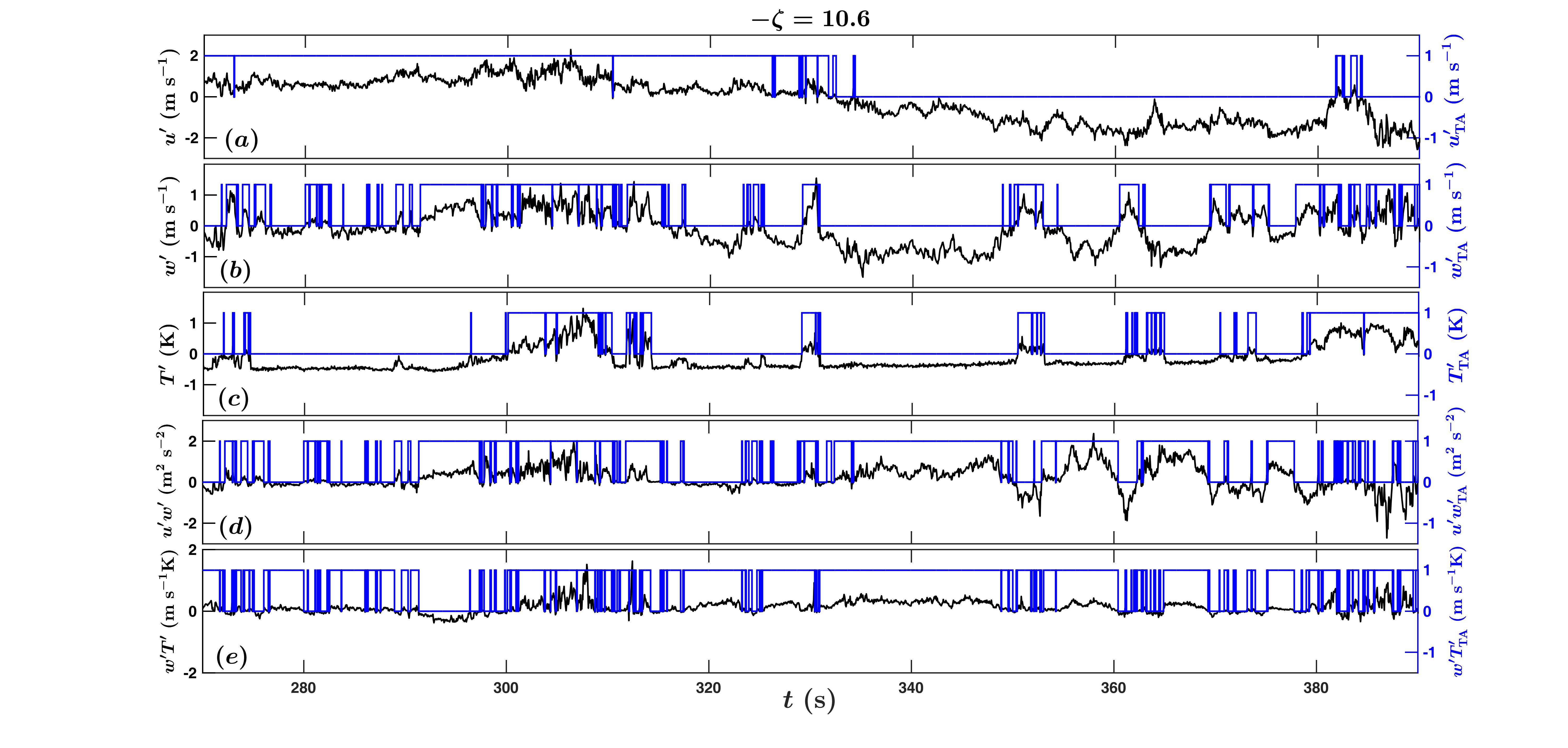}
  \caption{A 120-s long section of a time series of (a) $u^{\prime}$, (b) $w^{\prime}$, (c) $T^{\prime}$, (d) $u^{\prime}w^{\prime}$, and (e) $w^{\prime}T^{\prime}$ from a highly convective surface layer corresponding to a $-\zeta=$ 10.6 are shown. The actual values are displayed as solid black lines which correspond to the left hand side of the $y$ axis. Similarly, the associated telegraphic approximations (TA) are displayed as solid blue lines, which correspond to the right hand side of the $y$ axis.}
\label{fig:1}
\end{figure*}

A graphical demonstration of the persistence phenomenon is provided in figure \ref{fig:1}, where a 120-s long section of streamwise velocity fluctuations ($u^{\prime}$), vertical velocity fluctuations ($w^{\prime}$), temperature fluctuations ($T^{\prime}$), and their products representing the instantaneous streamwise momentum ($u^{\prime}w^{\prime}$) and heat fluxes ($w^{\prime}T^{\prime}$) are shown for a particular 30-min run, corresponding to $-\zeta=10.6$. The associated telegraphic approximations (TA) of these time series are presented at the right hand side of the $y$ axis in figure \ref{fig:1}, displayed in blue (see Section \ref{persistence} for additional details). A visual inspection of figure \ref{fig:1} suggests that the $w^{\prime}$ signals switch more often from the positive (negative) to the negative (positive) turbulent states, as compared to the $u^{\prime}$ or the $T^{\prime}$ signals. Nevertheless, the streamwise momentum ($u^{\prime}w^{\prime}$) and the heat flux ($w^{\prime}T^{\prime}$) signals involve both the combinations of the $w^{\prime}$ and $u^{\prime}$ or $w^{\prime}$ and $T^{\prime}$ signals. Therefore, the switching tendencies of these turbulent fluxes (figures \ref{fig:1}d and e) must be related to the persistence signatures of their component signals. This issue is discussed further in Section \ref{persistence}. 

As described by \citet{chowdhuri2020persistence}, the persistence time ($t_{p}$) is defined as the time up to which a fluctuating turbulent signal stays positive or negative before being switched to the other state. In addition to that, the associated probability density function (PDF) of $t_{p}$ describes its statistical characteristics, which in turn are related to the turbulent structures in a convective flow \citep{chowdhuri2020persistence}. We apply these same concepts in the present study to investigate the persistence behaviour of the instantaneous heat and momentum flux signals. Typically, we encounter in the order of 10$^{5}$ number of zero-crossings (the points where the TA series changes from 0 to 1 or 1 to 0) for the $u^{\prime}$, $w^{\prime}$, $T^{\prime}$, $u^{\prime}w^{\prime}$, and $w^{\prime}T^{\prime}$ signals for every six stability classes \citep{chowdhuri2020persistence}. Therefore, the persistence PDFs of the $u^{\prime}w^{\prime}$ and $w^{\prime}T^{\prime}$ signals for each of these six stability classes are constructed over these large number of ensemble events, to ensure their statistical robustness. Note that, the persistence PDFs are computed via logarithmic binning and subsequent transformation to the linear space using a change of variable, as illustrated by \citet{chowdhuri2020persistence}. In the following section, we discuss the properties of the persistence PDFs of the streamwise momentum and the heat flux signals, corresponding to these six stability classes.

\section{Results and discussion}
\label{results}
We begin by presenting the results of the persistence PDFs of the instantaneous momentum and heat flux signals ($u^{\prime}w^{\prime}$ and $w^{\prime}T^{\prime}$), including a comparison with the persistence of the component signals themselves ($u^{\prime}$, $w^{\prime}$, and $T^{\prime}$). Subsequently, we discuss the effect of separation into four different quadrants on the persistence PDFs of the flux signals. Furthermore, we explore the role of the intermittent flux events of different persistence time scales towards the heat and momentum transport efficiencies. In order to achieve such objective, we introduce a novel approach to separate the phases and amplitudes of the component signals associated with such events of various time scales. Additionally, we also provide plausible physical explanations of the obtained results during the course of our presentation. 

\subsection{Persistence PDFs of heat and momentum fluxes}
\label{persistence}
Prior to embarking on a detailed analysis regarding the persistence PDFs of the heat and momentum flux signals, perhaps it is prudent to establish a phenomenological connection between the persistence properties of the fluxes and their component signals. In order to derive such a heuristic relation, we turn our attention towards the telegraphic approximations (TA) of the turbulent signals. Specifically, we ask the question \emph{if the TA representation of the component signals are known, then what would be the equivalent TA representation of their product?}

\subsubsection{Association between the persistence time scales of the flux and its components}
\label{connection}
The TA representation of a turbulent signal can be expressed as,
\begin{equation}
(x^{\prime})_{\rm TA}=\frac{1}{2}(\frac{x^{\prime}(t)}{|x^{\prime}(t)|}+1),
\label{TA}
\end{equation}
where $x$ can be either $u$, $w$, or $T$. From the available measurements if we know the TA representation of the two signals, lets say $u^{\prime}$ and $w^{\prime}$, then the equivalent TA representation of their product $u^{\prime}w^{\prime}$ can be written as,
\begin{equation}
(u^{\prime}w^{\prime})_{\rm TA}=1-|u^{\prime}_{\rm TA}-w^{\prime}_{\rm TA}|.
\label{TA_1}
\end{equation}
Note that, the same expression is true for the product of the $w^{\prime}$ and $T^{\prime}$ signals as well. However, for the illustration purpose we will limit our discussion to the product of the $u^{\prime}$ and $w^{\prime}$ signals only, although the similar logic can be extended to any two signals. 

At a first glance the relationship provided in Eq. (\ref{TA_1}) may seem counter-intuitive, as one may naively expect the TA representation of the product of the two signals will be equal to the product of the TA's of the individual components. Nevertheless, expressing the TA representation of the product in such way will be incorrect. This is because, if both the signals are negative (individual TA's equal to 0) the products are positive and hence their TA representations are equal to 1. To demonstrate this, in figure S1 (supplementary material) we provide a typical example of the original TA approximated time series of $u^{\prime}w^{\prime}$ and $w^{\prime}T^{\prime}$ for a highly-convective stability, corresponding to $-\zeta=10.6$. We compare the original $(u^{\prime}w^{\prime})_{\rm TA}$ and $(w^{\prime}T^{\prime})_{\rm TA}$ with Eq. (\ref{TA_1}) and with the product $u^{\prime}_{\rm TA} \times w^{\prime}_{\rm TA}$. The result clearly shows that the expression $u^{\prime}_{\rm TA} \times w^{\prime}_{\rm TA}$ does not capture the original TA signatures of $(u^{\prime}w^{\prime})_{\rm TA}$ and $(w^{\prime}T^{\prime})_{\rm TA}$, whereas Eq. (\ref{TA_1}) detains that information perfectly well.

From the definition of persistence, the time scale $t_{p}$ of any fluctuating signal ($x^{\prime}$) can be expressed as,
\begin{equation}
t|(\Delta x^{\prime}_{\rm TA}=0) \implies t_{p},
\label{TA_3}
\end{equation}
where $\Delta x^{\prime}_{\rm TA}$ is the difference between the successive TA values of $x^{\prime}_{\rm TA}$. To connect the persistence time scales of the component signals with their product, we can transform Eq. (\ref{TA_1}) in the difference form as,
\begin{equation}
|\Delta (u^{\prime}w^{\prime})_{\rm TA}|=\Big||\Delta u^{\prime}_{\rm TA}|-|\Delta w^{\prime}_{\rm TA}| \Big|.
\label{TA_4}
\end{equation}
From Eq. (\ref{TA_4}), one can infer that the zero-crossings of $(u^{\prime}w^{\prime})_{\rm TA}$ will be located at those points where, $|\Delta u^{\prime}_{\rm TA}| \neq |\Delta w^{\prime}_{\rm TA}|$. Therefore, the persistence time scales of $u^{\prime}w^{\prime}$ will be related to the persistence time of that component signal which switches its states more rapidly than the other one. However, there is also an another alternate scenario for a hypothetical case where the $u^{\prime}$ and $w^{\prime}$ signals are perfectly correlated with each other. In such scenario, $|\Delta u^{\prime}_{\rm TA}|$ will be equal to $|\Delta w^{\prime}_{\rm TA}|$ at all time scales, which will result in an infinite persistence of their product $u^{\prime}w^{\prime}$.

Nevertheless, for all practical purposes, we can anticipate that the persistence PDFs of the products of the turbulent signals will be closer to the persistence PDFs of the individual signals at small time scales. The reason is the probability of occurrence of the events persisting for short times will not be much different for the $u^{\prime}$, $w^{\prime}$, and $u^{\prime}w^{\prime}$ signals, since all of them rapidly fluctuate. On the other hand, at large time scales, the persistence PDFs of the products are expected to be more closer to that signal which has a lower chance of encountering such long sequences. Note that, similar inference can also be drawn for the product between $w^{\prime}$ and $T^{\prime}$. Next, we describe how the insights gained from the aforementioned analysis compare with the observations of the persistence PDFs of heat and momentum fluxes in a convective ASL flow.  

\subsubsection{Features of the flux persistence PDFs}
\label{features}
\begin{figure}[h]
\centering
\hspace*{-0.7in}
\includegraphics[width=1.25\textwidth]{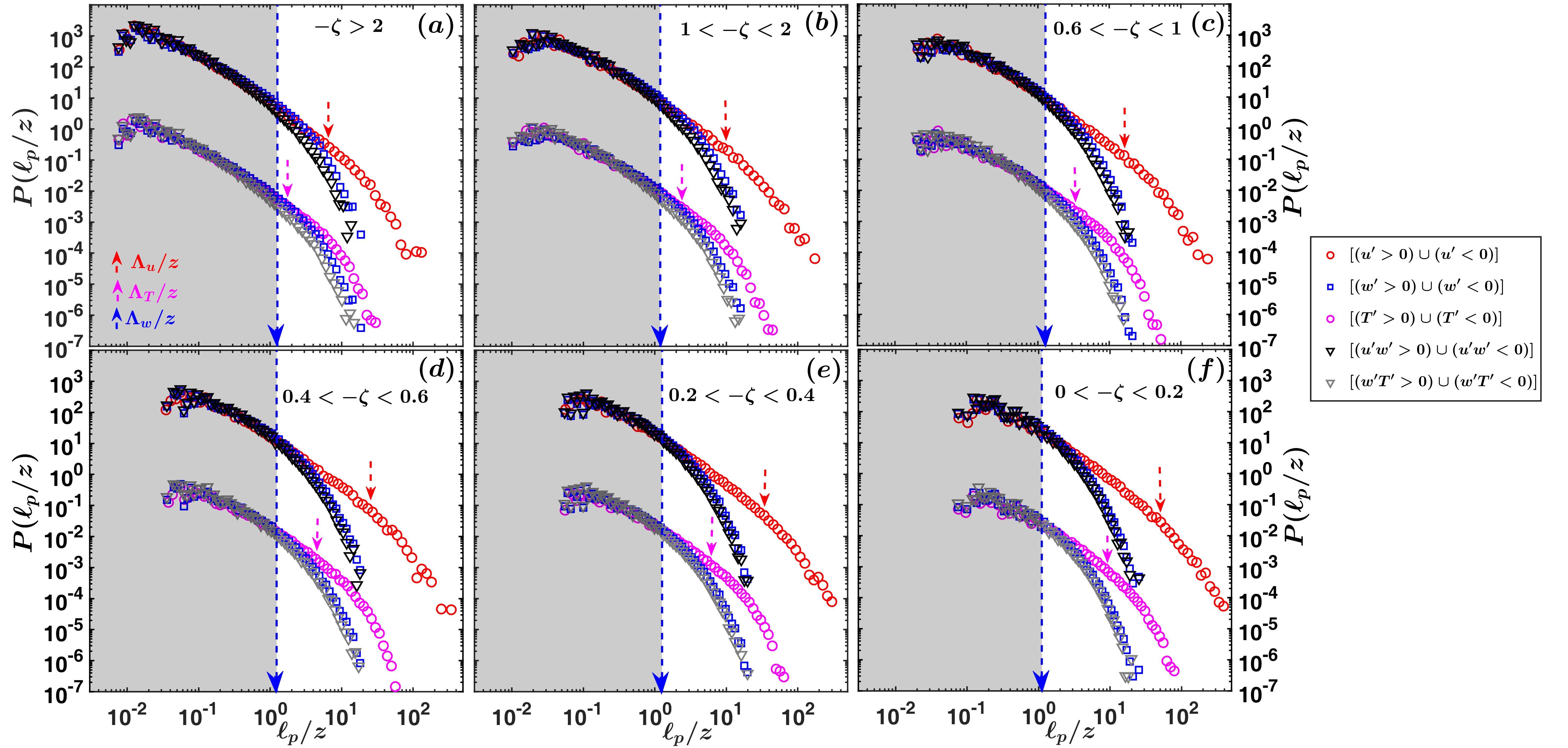}
  \caption{The persistence PDFs of the normalized sizes $\ell_{p}/z$ corresponding to the streamwise velocity fluctuations ($u^{\prime}$, red circles), temperature fluctuations ($T^{\prime}$, pink circles), vertical velocity fluctuations ($w^{\prime}$, blue squares), streamwise momentum flux ($u^{\prime}w^{\prime}$, black inverted triangles), and the kinematic heat flux ($w^{\prime}T^{\prime}$, grey inverted triangles) are shown separately for the six different stability classes. The persistence PDFs of $w^{\prime}$ are plotted twice, since both the momentum and heat flux signals involve the vertical velocity fluctuations. The panels corresponding to these six stability classes are arranged from the top-left to the bottom-right as, (a) $-\zeta>2$, (b) $1<-\zeta<2$, (c) $0.6<-\zeta<1$, (d) $0.4<-\zeta<0.6$, (e) $0.2<-\zeta<0.4$, and (f) $0<-\zeta<0.2$. Note that, the persistence PDFs for all the variables are computed after considering both the positive and negative fluctuations together. For visualization purpose, the persistence PDFs of $u^{\prime}$, $u^{\prime}w^{\prime}$, and the associated $w^{\prime}$ signals are shifted vertically upwards by two decades. The descriptions of the markers are provided in the legend, placed at the right-most side of the figure. The coloured arrows show the positions of the normalized integral length scales, corresponding to $w^{\prime}$ ($\Lambda_{w}/z$, blue arrows), $T^{\prime}$ ($\Lambda_{T}/z$, pink arrows), and $u^{\prime}$ ($\Lambda_{u}/z$, red arrows) signals. The regions on all the panels corresponding to $\ell_{p}/\Lambda_{w} \leq 1$ are marked in grey.}
\label{fig:2}
\end{figure}

Figure \ref{fig:2} shows the persistence PDFs of the instantaneous streamwise momentum ($u^{\prime}w^{\prime}$) and heat fluxes ($w^{\prime}T^{\prime}$), compared with their component signals ($u^{\prime}$ and $w^{\prime}$ or $w^{\prime}$ and $T^{\prime}$) for the six stability classes in a convective ASL flow. In figure \ref{fig:2}, the persistence time scales are converted to streamwise lengths ($t_{p}\overline{u}$, where $\overline{u}$ is the mean wind speed) by applying the Taylor's frozen turbulence hypothesis and scaled with $z$. Such scaling stems from the supposition that the eddies near the surface linearly grow with $z$ \citep{tennekes1972first,davidson2015turbulence}. Hereafter, we denote the converted streamwise lengths related to persistence as $\ell_{p}$. Apart from that, the coloured dashed arrows in figure \ref{fig:2} indicate the normalized integral length scales of $u^{\prime}$ ($\Lambda_{u}/z$, red dashed arrows), $w^{\prime}$ ($\Lambda_{w}/z$, blue dashed arrows), and $T^{\prime}$ signals ($\Lambda_{T}/z$, pink dashed arrows) computed from the exponential fits to the respective auto-correlation functions \citep [see their figure 5 for the auto-correlation plots]{chowdhuri2020persistence}. 

One can notice from figure \ref{fig:2} that, irrespective of the stability classes, for length scales smaller than the integral scale of $w^{\prime}$ ($\ell_{p} \leq \Lambda_{w}$), the persistence PDFs of momentum or heat nearly collapse on the persistence PDFs of the individual component signals such as, $u^{\prime}$ and $w^{\prime}$ or $T^{\prime}$ and $w^{\prime}$. However, for $\ell_{p}>\Lambda_{w}$ the persistence PDFs of the turbulent fluxes more closely follow the PDFs associated with the $w^{\prime}$ signal. The reason behind this observed phenomenon is twofold and in accord with the inferences drawn from Eq. (\ref{TA_4}). First, the integral scales of the $u^{\prime}$ and $T^{\prime}$ signals remain significantly larger than the $w^{\prime}$ signal. Second, \citet{chowdhuri2020persistence} have demonstrated that the persistence PDFs follow a power-law distribution up to the scales comparable with the integral scale before they start to drop off exponentially. Combining these two rationales, it is imperative that at larger time scales, greater than $\Lambda_{w}$, the persistence PDFs of $w^{\prime}$ would fall off faster than the $u^{\prime}$ or $T^{\prime}$ signals, given $\Lambda_{w}<\Lambda_{u,T}$. As a consequence, the flux persistence PDFs of $u^{\prime}w^{\prime}$ and $w^{\prime}T^{\prime}$ follow the same PDFs of the $w^{\prime}$ signals at scales larger than $\Lambda_{w}$, whereas nearly matching with both the individual signals for the scales $\ell_{p} \leq \Lambda_{w}$.

In summary, this analysis show that the persistence PDFs of the turbulent fluxes have two distinct characteristics which can be separated based on the integral scale of $w^{\prime}$. At scales smaller than $\Lambda_{w}$, the persistence PDFs of $u^{\prime}w^{\prime}$ and $w^{\prime}T^{\prime}$ are influenced by the power-law regions of their component PDFs. Contrarily, at scales larger than $\Lambda_{w}$, the persistence PDFs of the turbulent fluxes follow the $w^{\prime}$ signal quite well. This is because the persistence PDFs of $w^{\prime}$ decrease faster than the $u^{\prime}$ or $T^{\prime}$ signals.

To explore this further, we note that the turbulent transports of heat and momentum are accomplished through events from four different quadrants, namely the down-gradient and counter-gradient quadrants \citep{wallace2016quadrant}. In a convective ASL, the transport efficiencies of the heat and momentum are intimately related to how the total fluxes are partitioned between the down-gradient and counter-gradient quadrants \citep{li2011coherent}. Therefore, as a first step to connect the persistence behaviour of the fluxes at different scales with their transport characteristics, it is important to decompose the heat and momentum flux signals into four different quadrants. We present the results associated with this aspect in Section \ref{quadrants}.  

\subsubsection{The quadrant characterization of the flux persistence PDFs}
\label{quadrants}
\begin{figure}[h]
\centering
\hspace*{-0.7in}
\includegraphics[width=1.25\textwidth]{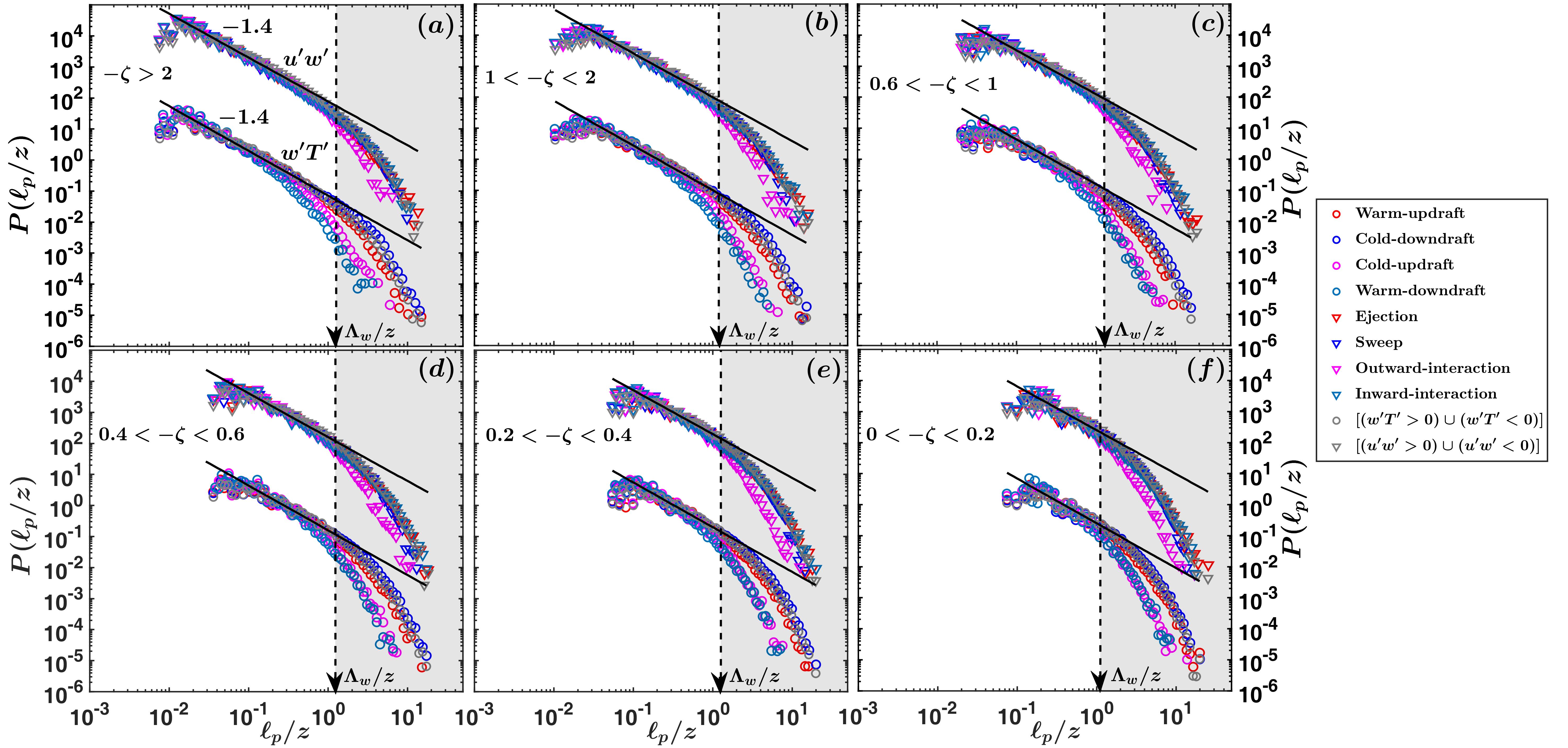}
  \caption{Persistence PDFs of the streamwise momentum ($u^{\prime}w^{\prime}$, inverted triangles), and the heat flux ($w^{\prime}T^{\prime}$, circles) signals, considered separately for the four different quadrants. The panels corresponding to the six stability classes are arranged from the top-left to the bottom-right as, (a) $-\zeta>2$, (b) $1<-\zeta<2$, (c) $0.6<-\zeta<1$, (d) $0.4<-\zeta<0.6$, (e) $0.2<-\zeta<0.4$, and (f) $0<-\zeta<0.2$.  For the visualization purpose, the persistence PDFs of the $u^{\prime}w^{\prime}$ signals associated with all the four different quadrants are shifted vertically upwards by three decades. The thick black lines show the power-laws with their respective slopes being mentioned on the panel (a). The grey inverted triangles and circles denote the $u^{\prime}w^{\prime}$ and $w^{\prime}T^{\prime}$ persistence PDFs, computed after considering both the positive and negative values together. The descriptions of the markers are provided in the legend, placed at the right-most side of the figure. The vertical dashed arrows in black indicate the position $\ell_{p}/\Lambda_{w}=1$. The regions on all the panels corresponding to $\ell_{p}/\Lambda_{w} > 1$ are marked in grey.}
\label{fig:3}
\end{figure}

Figure \ref{fig:3} displays the persistence PDFs of $u^{\prime}w^{\prime}$ and $w^{\prime}T^{\prime}$ signals decomposed into four different quadrants. Additionally, we also compare the persistence PDFs from four quadrants with the total persistence of the heat and momentum signals (see the respective grey circles and inverted triangles in figure \ref{fig:3}). Such comparison is required to identify at what scales which quadrant events dominate the persistence behaviour of the flux signals. 

An apparent observation from figure \ref{fig:3} is, for the $u^{\prime}w^{\prime}$ signal, the persistence PDFs corresponding to all the four quadrants and their total are in excellent agreement with each other, irrespective of the stability conditions. Nevertheless, there is a faint indication that the $u^{\prime}w^{\prime}$ persistence PDFs concurrent with the outward-interaction quadrants slightly deviate from the rest of the three quadrants. On the other hand, for the $w^{\prime}T^{\prime}$ signal, approximately at scales $\ell_{p}>\Lambda_{w}$, the persistence PDFs corresponding to the down-gradient and counter-gradient quadrants remain largely separated from each other. To be precise, for the highly-convective stability class ($-\zeta>2$), the quadrant partition occurs at scales $\ell_{p} \approx 0.3 \Lambda_{w}$, instead of at $\Lambda_{w}$. However, with the decrease in $-\zeta$ the threshold indeed becomes closer to $\Lambda_{w}$. Inspite of such dependency, the threshold at $\Lambda_{w}$ serves well to differentiate the transport characteristics between the large and small scale flux events (see Section \ref{Degree_of_organization}). More importantly, the reason behind the quadrant separation for the heat flux events is tied to the fact that the probabilities of encountering long sequences of counter-gradient activities remain significantly low. Interestingly, the discrepancy between the quadrants is quite prominent for the highly-convective stability (figure \ref{fig:3}a), but becomes inconspicuous in the near-neutral conditions (figure \ref{fig:3}f). Moreover, it is also noticeable that irrespective of stability the total persistence PDFs of the $w^{\prime}T^{\prime}$ signals are in accord with the down-gradient quadrants.

Along with the aforementioned features, one can notice that up to the scale $\ell_{p} \leq \Lambda_{w}$, the persistence PDFs of the heat and momentum flux signals collapse onto each other following a power-law behaviour. Note that, such power-law signatures in the flux persistence PDFs are quite extensive for the highly-convective stability (figure \ref{fig:3}a). Therefore, the data from such stability conditions have been used for the estimation of its exponent. Since we plot the persistence PDFs in log-log plots where power-law appears as a straight line, the exponent is determined through linear regression for scales $0.01 \leq \ell_{p} \leq \Lambda_{w}$. For both the $u^{\prime}w^{\prime}$ and $w^{\prime}T^{\prime}$ signals, we obtain a best fit exponent of $-1.4$ with $R^{2}>0.95$. Moreover, to assess the effect of stability on the power-law exponent, the same power-law curves as obtained for the highly-convective stability are compared with the other five stability classes (figures \ref{fig:3}b to \ref{fig:3}f). Such comparison shows that, even though there is no discernible change in the slope of the power-law, its extent gradually gets shorter as the ASL approaches the near-neutral stability (figures \ref{fig:3}a to \ref{fig:3}f). Since the measurements from the near-neutral stability class belong to the lowest three SLTEST levels, the shrinkage in the power-law regime is related to insufficient sampling of the small scale eddies at 20-Hz sampling frequency \citep{chowdhuri2020persistence}. On the other hand, at scales $\ell_{p}>\Lambda_{w}$, the deviation from the power-law behaviour becomes notable for both the flux persistence PDFs where they follow an exponential distribution, hallmark of a Poisson process \citep{chowdhuri2019revisiting}. 

In a nutshell, figure \ref{fig:3} elucidates the effects of quadrant separation on the flux persistence PDFs. An emergent result is the quadrant segregation does not have any appreciable effect on the persistence PDFs of the momentum flux ($u^{\prime}w^{\prime}$) signals at all the scales of motions. Conversely, for the heat flux ($w^{\prime}T^{\prime}$) signals, the quadrant partition only has significance at scales $\ell_{p}>\Lambda_{w}$. Physically, this difference in quadrant behaviour between the $u^{\prime}w^{\prime}$ and $w^{\prime}T^{\prime}$ signals is related to transport asymmetry associated with the down-gradient and counter-gradient quadrants. Such interpretation is feasible, since the contrast between the persistence PDFs of positive and negative values corresponding to any stochastic signal is related to the skewness of that signal, as demonstrated by \citet{chowdhuri2020persistence}. It is worth noting that, in a highly-convective ASL, the PDFs of the $w^{\prime}T^{\prime}$ signals are significantly asymmetric between the positive and negative values due to the non-Gaussian character of the temperature fluctuations. On the contrary, the PDFs of the $u^{\prime}w^{\prime}$ signals are more symmetric because both the streamwise and vertical velocity fluctuations are close-to-Gaussian in nature \citep [see their figure 5]{chowdhuri2019revisiting}. Additionally, for scales $\ell_{p}>\Lambda_{w}$, both the flux persistence PDFs follow a Poisson distribution. Previous studies have surmised such phenomenon as a signature of random deformation of the coherent structures due to the presence of the ground \citep{cava2012role,chowdhuri2020persistence}. 

Nonetheless, for scales $\ell_{p} \leq \Lambda_{w}$, the persistence PDFs of both the fluxes collapse in a power-law fashion. The exponents of this power-law remain same for the $u^{\prime}w^{\prime}$ and $w^{\prime}T^{\prime}$ signals, being equal to $-1.4$. The analysis of \citet{chowdhuri2020persistence} has revealed that the power-law regimes of the persistence PDFs for the velocity and temperature signals are related to the self-similar Richardson cascade which constitute the inertial subrange of the turbulence spectrum. Besides, \citet{chowdhuri2020persistence} reported the respective power-law exponents to be equal to $-1.6$, $-1.25$, and $-1.4$ for the $u^{\prime}$, $w^{\prime}$, and $T^{\prime}$ signals. They linked the difference in the exponents to the disparity in the small-scale intermittency, expressed through the framework of self-organized criticality. Interestingly, the power-law exponents corresponding to the flux signals are $-1.4$ irrespective of their type and coincidentally being equal to the same for the $T^{\prime}$ signal. Further explanation behind this coincidence is elusive at present. 

Hitherto, from the detailed discussion of the flux persistence PDFs, one could sense that the persistence properties of the momentum and heat flux signals exhibit significantly different behaviour for scales smaller or larger than the integral scale of the vertical velocity. Since $\Lambda_{w}$ is of the same order as $z$ \citep{chowdhuri2020persistence}, such discrepancy reflects distinct attributes of turbulent transport associated with the detached and attached eddies, characterized by length scales smaller and larger than $z$ \citep{marusic2019attached}. However, it is important to remember that the persistence analysis only provides information on the distribution of time or length scales of the flux events but not on the flux values themselves \citep{bershadskii2004clusterization}. Therefore, such analysis alone is insufficient to deduce the effects of the flux events, distinguished by their length scales greater or lesser than $\Lambda_{w}$, on the heat and momentum transport efficiencies. In Section \ref{flux_vectors}, we introduce a methodology where the transport characteristics are studied separately for the heat and momentum flux events with length scales $\ell_{p}>\Lambda_{w}$ and $\ell_{p} \leq \Lambda_{w}$. 

\subsection{Polar representation of the quadrant planes}
\label{flux_vectors}
In general, the heat and momentum transport characteristics allied to the turbulent motions are expressed through quadrant analysis, where each point plotted on the $u^{\prime}$-$w^{\prime}$ or $T^{\prime}$-$w^{\prime}$ quadrant plane corresponds to a certain flux event. An example of such quadrant plot is provided in figure S2 (supplementary material) to explain the concepts. The usual practice in this analysis is to report the momentum or heat flux fractions and time fractions from each quadrant and assess the relative importance of the various turbulent motions, associated with different flow structures \citep{wallace2016quadrant}. Additionally, it is also worthwhile to mention that, the transport efficiencies of the heat and momentum are intimately linked to the partition of the fluxes among these four different quadrants \citep{li2011coherent,chowdhuri2019evaluation}.

However, on a rudimentary level, the existence of the heat and momentum flux depends upon the strength of the coupling between the two turbulent signals as described by the governing Navier-Stokes equations. In order to understand the cause of the coupling, it is useful to envisage the quadrant plane as a phase space by drawing analogies with non-linear dynamical systems. From this perspective, each point in the quadrant plane is related to a particular state of flux generation, designated with two parameters which are the phase angles and amplitudes. To accomplish such an objective, the polar representation of the quadrant plane is a well-suited approach \citep{mahrt1984heat}. One should note that, in this approach the amplitudes are not the intensities of the flux themselves, since the flux values are a combination of both the phases and amplitudes (see Eq. (\ref{flux_value_pre})). 

\begin{figure}[h]
\centering
\hspace*{-1.5in}
\includegraphics[width=1.3\textwidth]{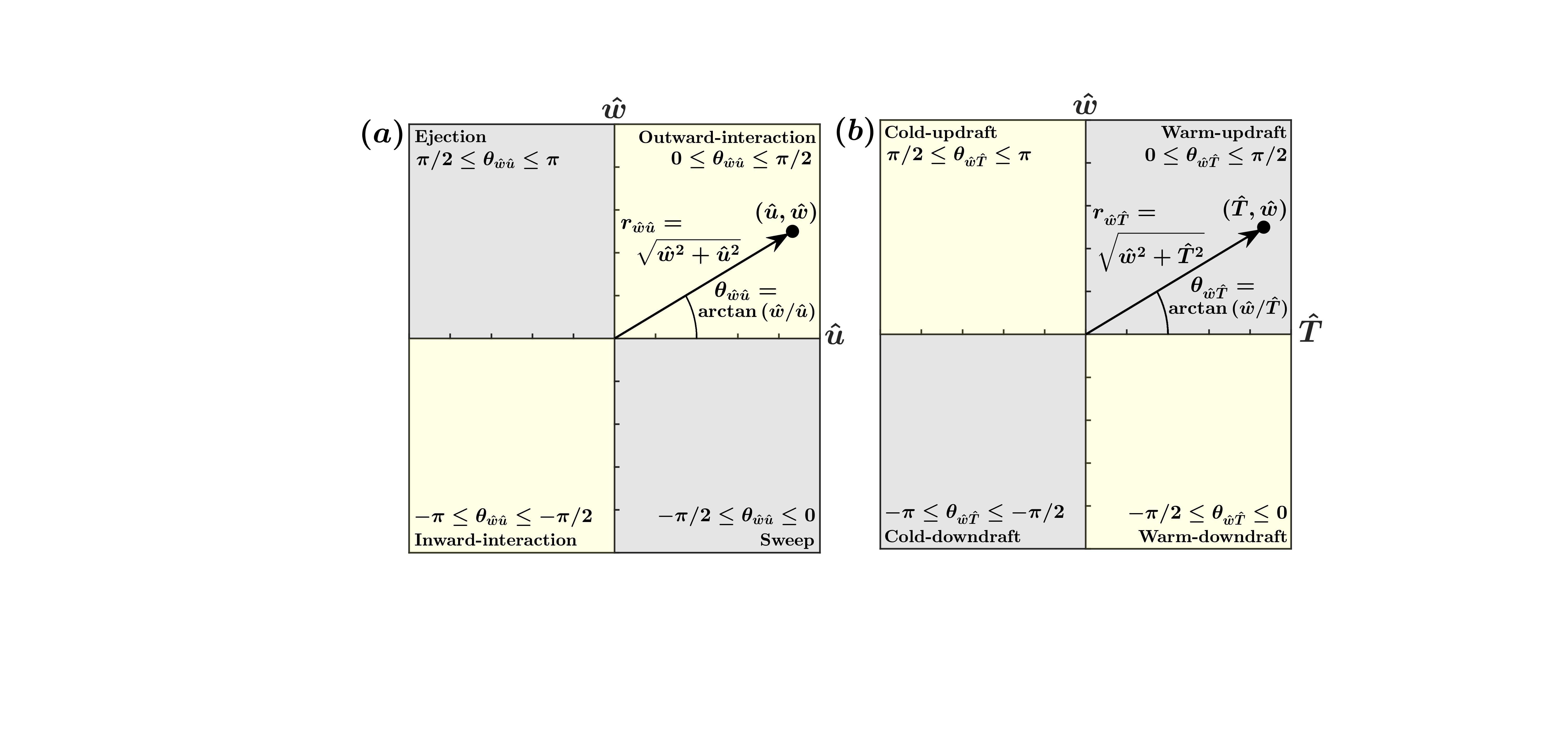}
\vspace{-2cm}
  \caption{The graphical illustrations of the (a) $u^{\prime}$-$w^{\prime}$ and (b) $T^{\prime}$-$w^{\prime}$ quadrant planes are provided. The fluctuations in $u$, $w$, and $T$ are normalized by their respective standard deviations ($\sigma_{u}$, $\sigma_{w}$, and $\sigma_{T}$) and denoted as $\hat{x}$ ($\hat{x}=x^{\prime}/\sigma_{x}$, where $x$ can be $u$, $w$, or $T$). The four different quadrants are marked on the respective panels. In a polar co-ordinate representation, each point on the quadrant plane (\{$\hat{u},\hat{w}$\} or \{$\hat{T},\hat{w}$\}) can be described by their amplitudes ($r_{\hat{w}\hat{u}}$ or $r_{\hat{w}\hat{T}}$) and the phase angles ($\theta_{\hat{w}\hat{u}}$ or $\theta_{\hat{w}\hat{T}}$). The phase angles can take values between $-\pi$ to $\pi$ depending on the position of the quadrants, as displayed on the panels. The grey shaded regions in the quadrant planes denote the down-gradient quadrants, whereas the yellow shaded regions denote the counter-gradient quadrants.}
\label{fig:4}
\end{figure}

To explain such approach, figure \ref{fig:4} graphically illustrates the concept of quadrant planes of $u^{\prime}$-$w^{\prime}$ and $T^{\prime}$-$w^{\prime}$ from the perspective of a polar reference frame. Following the standard practice, the $u^{\prime}$, $w^{\prime}$, and $T^{\prime}$ signals in figure \ref{fig:4} are normalized by their respective standard deviations ($\sigma_{u}$, $\sigma_{w}$, and $\sigma_{T}$) and denoted as $\hat{x}$ ($\hat{x}=x^{\prime}/\sigma_{x}$, where $x$ can be $u$, $w$, or $T$). Such normalization is necessary since it removes the issue regarding the difference in the units, while computing the phase angles and amplitudes (defined below). In the polar reference frame, each point on the $\hat{x}$-$\hat{w}$ (where $x$ can be $u$ or $T$) quadrant plane can be associated with an amplitude $r_{\hat{w}\hat{x}}$ and phase angle $\theta_{\hat{w}\hat{x}}$, expressed as,
\begin{align}
\theta_{\hat{w}\hat{x}}=\arctan{(\hat{w}/\hat{x})}\\
r_{\hat{w}\hat{x}}=\sqrt{\hat{w}^2+\hat{x}^2}
\label{flux_vector}.
\end{align}
The lengths of the phase vectors (shown as arrows) on figure \ref{fig:4} indicate the values of $r_{\hat{w}\hat{x}}$, whereas $\theta_{\hat{w}\hat{x}}$ are the angles subtended by these vectors with the $x$ axes. The values of the phase angles vary between $-\pi$ to $\pi$ and their ranges are related to the four different quadrants as demonstrated in table \ref{tab:1}. For identification purposes, the $\theta_{\hat{w}\hat{x}}$ ranges which denote the down-gradient (counter-gradient) quadrants are marked as grey (yellow) shaded regions in figure \ref{fig:4}. 
\begin{table}[h]
  \caption{The four quadrants of $\hat{u}$-$\hat{w}$ and $\hat{T}$-$\hat{w}$ and the associated distribution of phase angles in a convective ASL}
  \begin{center}
\def~{\hphantom{0}}
  \begin{tabular}{lcccc}
  \toprule
Phase angle & $\hat{u}$-$\hat{w}$ quadrant & Quadrant name & $\hat{T}$-$\hat{w}$ quadrant & Quadrant name \\
\\
$0 \leq \theta_{\hat{w}\hat{u}}, \theta_{\hat{w}\hat{T}} \leq \pi/2$ & $\hat{u}>$ 0, $\hat{w}>$ 0 & Outward-interaction & $\hat{w}>$ 0, $\hat{T}>$ 0 & Warm-updraft\\
\\
$\pi/2 \leq \theta_{\hat{w}\hat{u}}, \theta_{\hat{w}\hat{T}} \leq \pi$ & $\hat{u}<$ 0, $\hat{w}>$ 0 & Ejection & $\hat{w}>$ 0, $\hat{T}<$ 0 & Cold-updraft \\
\\
$-\pi \leq \theta_{\hat{w}\hat{u}}, \theta_{\hat{w}\hat{T}} \leq -\pi/2$ & $\hat{u}<$ 0, $\hat{w}<$ 0 & Inward-interaction & $\hat{w}<$ 0, $\hat{T}<$ 0 & Cold-downdraft\\
\\
$-\pi/2 \leq \theta_{\hat{w}\hat{u}}, \theta_{\hat{w}\hat{T}} \leq 0$ & $\hat{u}>$ 0, $\hat{w}<$ 0 & Sweep & $\hat{w}<$ 0, $\hat{T}>$ 0 & Warm-downdraft\\
\bottomrule
  \end{tabular}
 \label{tab:1}
  \end{center}
\end{table}
In the polar co-ordinate system, the normalized instantaneous flux ($\hat{u}\hat{w}$ or $\hat{w}\hat{T}$) associated with each point is expressed as $(r_{\hat{w}\hat{x}}^{2}\sin{2\theta_{\hat{w}\hat{x}}})/2$. This is because,
\begin{equation}
\hat{x}\hat{w}=r_{\hat{w}\hat{x}} \cos{(\theta_{\hat{w}\hat{x}})} \times r_{\hat{w}\hat{x}} \sin{(\theta_{\hat{w}\hat{x}})} \implies \frac{1}{2} r_{\hat{w}\hat{x}}^{2} \sin{(2 \theta_{\hat{w}\hat{x}})},
\label{flux_value_pre}
\end{equation}
given $\sin{(2 \theta_{\hat{w}\hat{x}})}=2\sin{(\theta_{\hat{w}\hat{x}})}\cos{(\theta_{\hat{w}\hat{x}})}$ and $x$ can be either $u$ or $T$. 

Subsequently, the averaged normalized flux over all the four quadrants can be written as,
\begin{equation}
\overline{\hat{x}\hat{w}}=\frac{1}{2}\overline{r_{\hat{w}\hat{x}}^{2}\sin{2\theta_{\hat{w}\hat{x}}}}.
\label{flux_value}
\end{equation}
The quantity on the left hand side of Eq. (\ref{flux_value}) is the correlation coefficient ($R_{wx}$) which indicates the transport efficiency of the momentum or the heat fluxes. Apart from that, since $r_{\hat{w}\hat{x}}^{2}$ is a positive definite quantity, the distribution of the fluxes among the four quadrants is primarily decided by the phase angle $\theta_{\hat{w}\hat{x}}$. Therefore, to simplify the expression in Eq. (\ref{flux_value}) we may replace the $r_{\hat{w}\hat{x}}^{2}$ values of every point in the quadrant plane by a single averaged value, $\overline{r_{\hat{w}\hat{x}}^{2}}$. Note that, 
\begin{equation}
\overline{r_{\hat{w}\hat{x}}^{2}}=\overline{{\big({\frac{x^{\prime}}{\sigma_{x}}}\big)}^2+{\big({\frac{w^{\prime}}{\sigma_{w}}}\big)}^2} \implies \overline{{\big({\frac{x^{\prime}}{\sigma_{x}}}\big)}^2}+\overline{{\big({\frac{w^{\prime}}{\sigma_{w}}}\big)}^2}.
\label{flux_value_amp}
\end{equation}
Since $\overline{{({{x^{\prime}}/{\sigma_{x}}})}^2}$ and $\overline{{({{w^{\prime}}/{\sigma_{w}}})}^2}$ are both equal to 1, from Eq. (\ref{flux_value_amp}) it indicates $\overline{r_{\hat{w}\hat{x}}^{2}}=2$. Consequently, this enables us to simplify Eq. (\ref{flux_value}) as,
\begin{equation}
R_{wx}=\frac{\overline{r_{\hat{w}\hat{x}}^{2}}}{2} \times \overline{\sin{2\theta_{\hat{w}\hat{x}}}} \implies \overline{\sin{2\theta_{\hat{w}\hat{x}}}}
\label{flux_value_1}
\end{equation}
by substituting $\overline{r_{\hat{w}\hat{x}}^{2}}=2$. If the two signals are strictly phase-locked, then the phase angle $\theta_{\hat{w}\hat{x}}$ can take only two values which are $\pm \ \pi/4$ and $\pm \ 3\pi/4$. By replacing those in Eq. (\ref{flux_value_1}), one can deduce the correlation coefficients ($R_{wx}$) for such occasions would be perfectly $\pm \ 1$. 

To scrutinize this further, from the theory of probability one can modify Eq. (\ref{flux_value_1}) as,
\begin{equation}
R_{wx}=\int_{-\pi}^{\pi} P(\theta_{\hat{w}\hat{x}})\sin{(2\theta_{\hat{w}\hat{x}})} \ d\theta_{\hat{w}\hat{x}},
\label{flux_value_2}
\end{equation}
where $P(\theta_{\hat{w}\hat{x}})$ is the PDF of the phase angle $\theta_{\hat{w}\hat{x}}$, with the constraint, 
\begin{equation}
\int_{-\pi}^{\pi} P(\theta_{\hat{w}\hat{x}}) \ d\theta_{\hat{w}\hat{x}}=1.
\label{flux_pdf}
\end{equation}
Let us consider an another case, where the $\theta_{\hat{w}\hat{x}}$ values are uniformly distributed over the quadrant plane, such that $P(\theta_{\hat{w}\hat{x}})=1/(2\pi)$. This scenario would arise if the phase vectors were oriented in random directions with no phase-locking whatsoever. In that situation, Eq. (\ref{flux_value_2}) would be simplified as,
\begin{equation}
R_{wx}=\frac{1}{2\pi} \int_{-\pi}^{\pi} \sin{(2\theta_{\hat{w}\hat{x}})} \ d\theta_{\hat{w}\hat{x}}.
\label{flux_value_2_1}
\end{equation}
Since $\sin{(2 \theta_{\hat{w}\hat{x}})}$ is an odd function, from elemental calculus we know that the output of the integration in Eq. (\ref{flux_value_2_1}) is 0. This implies a completely inefficient transport of the turbulent fluxes. As a result, the heat and momentum transport efficiency can be judged by investigating the departure of the respective phase angle PDFs from a uniform distribution. An implicit assumption in such argument is, the role played by the amplitudes can be considered to be independent of the phase angles.  

\begin{figure}[h]
\centering
\hspace*{-1.5in}
\includegraphics[width=1.3\textwidth]{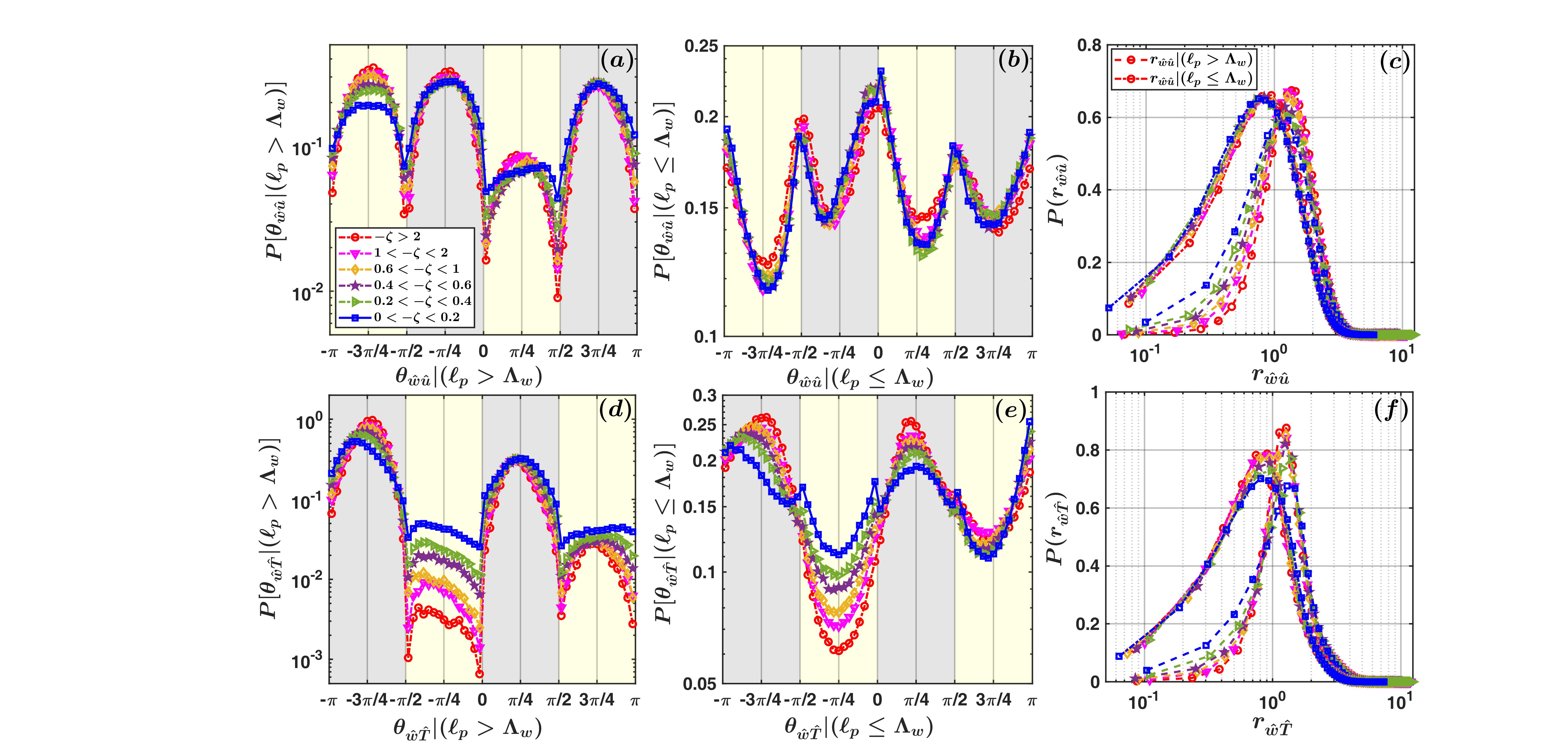}
  \caption{The PDFs of $\theta_{\hat{w}\hat{u}}$ are shown for the momentum flux events of length scales (a) $\ell_{p} > \Lambda_{w}$ or (b) $\ell_{p} \leq \Lambda_{w}$. Similarly, the PDFs of $\theta_{\hat{w}\hat{T}}$ are shown for the heat flux events corresponding to (d) $\ell_{p} > \Lambda_{w}$ and (e) $\ell_{p} \leq \Lambda_{w}$. The associated PDFs of the amplitudes are shown in panels (c) and (f) for $r_{\hat{w}\hat{u}}$ and $r_{\hat{w}\hat{T}}$ respectively. The dash-dotted and the dashed lines in the panels (c) and (f) describe the respective momentum or heat flux events with length scales $\ell_{p} \leq \Lambda_{w}$ and $\ell_{p} > \Lambda_{w}$ (see the legend in panel (c)). The description of the various markers indicating the six different stability classes is provided in the legend of panel (a). The grey and yellow shaded regions correspond to the down-gradient and counter-gradient quadrants respectively, as displayed in figure \ref{fig:4}.}
\label{fig:5}
\end{figure}

The aforementioned method based on the polar representation of the quadrant planes offers a unique opportunity to dissect the role of phase coupling towards the transport efficiencies of heat and momentum. It is therefore of practical interest to investigate this aspect corresponding to the flux events with length scales larger or smaller than $\Lambda_{w}$. The intention behind such separation criterion emerges from the flux persistence PDFs as presented in figures \ref{fig:2} and \ref{fig:3}. To briefly recapitulate, at length scales smaller than the integral scale of $w^{\prime}$, the persistence PDFs of the heat and momentum flux events from any quadrant follow an identical power-law distribution. Such a phenomenon demonstrates the effect of the detached eddies from the inertial subrange of the turbulence spectrum. On the contrary, at length scales larger than $\Lambda_{w}$ (reminiscent of the attached eddies), the persistence PDFs of the heat flux events are clearly dominated by the down-gradient motions and follow an exponential distribution. However, for the momentum flux events, even at these large length scales, the quadrant effect is negligible on their persistence PDFs, albeit the nature of the distribution remains exponential. To get further insights, in figure \ref{fig:5}, we present results to study the distributions of the phase angles and amplitudes in relation to the flux events, conditionally sampled based on the threshold $\ell_{p}=\Lambda_{w}$.

Figure \ref{fig:5} shows the PDFs of the phase angles ($\theta_{\hat{w}\hat{x}}$) and amplitudes ($r_{\hat{w}\hat{x}}$) associated with the small and large scale flux events, delineated accordingly with respect to their persistence time scales $\ell_{p}/\Lambda_{w} \leq 1$ and $\ell_{p}/\Lambda_{w}>1$. Note that, both of these PDFs are computed by collecting all the points which reside within the ensemble of flux events from a particular stability class and persisting for times as prescribed before. Hereafter, we denote these conditional PDFs of the phase angles as $P[\theta_{\hat{w}\hat{x}} \vert (\ell_{p} \leq \Lambda_{w})]$ and $P[\theta_{\hat{w}\hat{x}} \vert (\ell_{p} > \Lambda_{w})]$ respectively. Additionally, $P[\theta_{\hat{w}\hat{x}} \vert (\ell_{p} \leq \Lambda_{w})]$ and $P[\theta_{\hat{w}\hat{x}} \vert (\ell_{p} > \Lambda_{w})]$ individually satisfy the constraint in Eq. (\ref{flux_pdf}), so that when integrated over the limit $-\pi$ to $\pi$ the result is unity. Other than that, the grey and yellow shaded regions in figures \ref{fig:5}a, b, d, and e identify the ranges of the phase angles which correspond to the regions ascertained to the down-gradient and counter-gradient quadrants respectively (see table \ref{tab:1} and figure \ref{fig:4}). One may bear in mind that the areas occupied by the phase angle PDFs in these shaded regions are proportional to the fraction of time spent by the signal in that quadrant state. This is because,
\begin{equation}
(T_{f})_{\rm X}=\int_{-\pi}^{\pi} P(\theta_{\hat{w}\hat{x}}) I_{\rm X}(\theta_{\hat{w}\hat{x}}) \ d\theta_{\hat{w}\hat{x}},
\label{T_f}
\end{equation}
where $(T_{f})_{\rm X}$ is time fraction spent in quadrant X (X could be any one of the four quadrants) and $I_{\rm X}(\theta_{\hat{w}\hat{x}})$ is an identity function which is unity when $\theta_{\hat{w}\hat{x}}$ lies within quadrant X or zero elsewise. It is noteworthy that in a hypothetical framework, if the two signals are completely phase-locked (phase difference of $0$ or $\pi$) the PDFs of the phase angles would be a superposition of two Dirac-delta functions at $\theta_{\hat{w}\hat{x}}$ values of $\pm \ \pi/4$ and $\pm \ 3\pi/4$. However, for all practical purposes such case is not possible since it would imply an infinite persistence of the flux signals as they would never cross the zero. That being said, we thus expect the phase angle PDFs to be distributed in a particular way over all the possible $\theta_{\hat{w}\hat{x}}$. 

From figure \ref{fig:5}a, the PDFs of the phase angles associated with the large scale momentum flux events display a prominent peak related to the ejection motions (the grey shaded region on the right), irrespective of the stability classes. However, for the negative values of the phase angles, the PDFs have two distinct peaks in the grey and yellow shaded regions with approximately similar values. From Eq. (\ref{T_f}), this suggests that for the large scale momentum flux events, there is almost an equal tendency for the phase vectors to either reside within the sweep or within the inward-interaction quadrants. Nevertheless, a minute change in such behaviour is observed for the near-neutral stability class. In those conditions, the PDF peak values within the sweep quadrants slightly exceed the values associated with the inward-interaction quadrants. On the other hand, for all the stability classes, the PDFs of the phase angles associated with the large scale heat flux events exhibit a bi-modal behaviour with two distinguished peaks corresponding to the warm-updraft and cold-downdraft motions (the right and left grey shaded regions in figure \ref{fig:5}d). Be that as it may, with the decrease in the thermal stratification, the PDF values of $\theta_{\hat{w}\hat{T}}\vert (\ell_{p} > \Lambda_{w})$ corresponding to the counter-gradient quadrants (yellow shaded regions in figure \ref{fig:5}d) show considerable increase. All of these features of the phase angle PDFs associated with the large scale heat and momentum flux events are in sync with the findings of \citet{haugen1971experimental}. They observed that in a convective ASL flow, the heat flux signals mainly exhibited large persistent positive events, whereas for the momentum flux long sequences of both negative and positive activities were common. 

From figure \ref{fig:5}, one could further notice that the shapes of the PDFs of the phase angles differ significantly between the large ($\ell_{p} > \Lambda_{w}$) and small ($\ell_{p} \leq \Lambda_{w}$) scale flux events (figures \ref{fig:5}a, b, d, and e). For the small scale momentum flux events, an immediate observation is the PDFs of the phase angles is nearly an inverted version of the PDFs associated with the large scale momentum flux events (figure \ref{fig:5}b). In figure \ref{fig:5}b, the maximum values of the PDFs for all the stability classes are located at around 0, with troughs replacing the peaks at approximately the same positions as in figure \ref{fig:5}a. Conversely, for the small scale heat flux events (figure \ref{fig:5}e), the behaviour of the phase angle PDFs is not completely opposite to that of the large scale events (figure \ref{fig:5}d), but the distinction between the two modes is definitely more vague.

In addition to the phase angles, we can also investigate the amplitude PDFs ($r_{\hat{w}\hat{x}}$) for both large and small scale heat and momentum flux events (figures \ref{fig:5}c and f). Such information would be important to evaluate whether there are any specific amplitudes which occur most often. From figures \ref{fig:5}c and f, we note that the PDFs of the amplitudes are uni-modal in character, with a shift in their peak positions for the large scale heat or momentum flux events (approximately from 0.8 to 1.2). Apart from that, the PDFs of the amplitudes collapse sufficiently well for all the stability classes, corresponding to the small scale flux events. On the other hand, for the large scale flux events, such collapse is relatively poor for the small values of the amplitudes, located to the left of the peak position.

In summary, the results obtained so far point out that the statistical characteristics of the phase angles associated with the heat and momentum flux events remain significantly different from each other. Inevitably, this effect is more clearly visible for the flux events with $\ell_{p}$ larger than the integral scale of the vertical velocity. To connect the behaviour of the phase angle PDFs with the transport efficiencies, from Eq. (\ref{flux_value}) we can write,
\begin{equation}
R_{wx}=\frac{1}{2} \Big[\overline {(r_{\hat{w}\hat{x}}^{2}\sin{2\theta_{\hat{w}\hat{x}}}) \vert (\ell_{p} > \Lambda_{w})+ (r_{\hat{w}\hat{x}}^{2}\sin{2\theta_{\hat{w}\hat{x}}})\vert(\ell_{p} \leq \Lambda_{w})} \Big],
\label{flux_value_3}
\end{equation}
where the quantity $(r_{\hat{w}\hat{x}}^{2}\sin{2\theta_{\hat{w}\hat{x}}})$ is divided between the events with scales $\ell_{p}>\Lambda_{w}$ and $\ell_{p} \leq \Lambda_{w}$. From Eq. (\ref{flux_value_2}), we can further expand Eq. (\ref{flux_value_3}) as,
\begin{equation}
\begin{split}
R_{wx} & =\frac{1}{2} \Big[ \langle {r_{\hat{w}\hat{x}}^{2}\vert(\ell_{p}>\Lambda_{w})} \rangle \int_{-\pi}^{\pi} P[\theta_{\hat{w}\hat{x}}\vert(\ell_{p}>\Lambda_{w})]\sin{[2 \theta_{\hat{w}\hat{x}}\vert(\ell_{p}>\Lambda_{w})]} \ d\theta_{\hat{w}\hat{x}} + \\
& \langle {r_{\hat{w}\hat{x}}^{2}\vert(\ell_{p} \leq \Lambda_{w})} \rangle \int_{-\pi}^{\pi} P[\theta_{\hat{w}\hat{x}}\vert(\ell_{p} \leq \Lambda_{w})]\sin{[2 \theta_{\hat{w}\hat{x}}\vert(\ell_{p} \leq \Lambda_{w})]} \ d\theta_{\hat{w}\hat{x}} \Big],
\end{split}
\label{flux_value_4}
\end{equation}
with the constraints,
\begin{align}
\int_{-\pi}^{\pi} P[\theta_{\hat{w}\hat{x}}\vert(\ell_{p}>\Lambda_{w})] \ d\theta_{\hat{w}\hat{x}}=1
\\
\int_{-\pi}^{\pi} P[\theta_{\hat{w}\hat{x}}\vert(\ell_{p} \leq \Lambda_{w})] \ d\theta_{\hat{w}\hat{x}}=1,
\label{flux_pdf_1}
\end{align}
after conditioning on the flux events based on their $\ell_{p}$ (as shown in figure \ref{fig:5}). Moreover, in Eq. (\ref{flux_value_4}) the angle brackets denote the most probable amplitudes over those events with length scales $\ell_{p}>\Lambda_{w}$ and $\ell_{p} \leq \Lambda_{w}$, respectively. As mentioned earlier, such ramifications rest on the assumption that the amplitude effects on the transport efficiency are independent from the phase angle distribution. We can thus replace those values as 1.2 and 0.8 respectively, obtained from the peak positions of the amplitude PDFs (see figures \ref{fig:5}c and f). Therefore, from Eqs. (\ref{flux_value_2}), (\ref{flux_value_2_1}), and (\ref{flux_value_4}) it is clear that, in order to evaluate the contribution of the flux events at different length scales to the heat and momentum transport efficiencies, one needs to estimate the departure of $P[\theta_{\hat{w}\hat{x}}\vert(\ell_{p}>\Lambda_{w})]$ and $P[\theta_{\hat{w}\hat{x}}\vert(\ell_{p} \leq \Lambda_{w})]$ from a uniform distribution. In Section \ref{Degree_of_organization} we present results to quantify such departure of the phase angle PDFs. 

\subsection{Phase angle PDFs and transport efficiency}
\label{Degree_of_organization}
\begin{figure}[h]
\centering
\hspace*{-1.5in}
\includegraphics[width=1.3\textwidth]{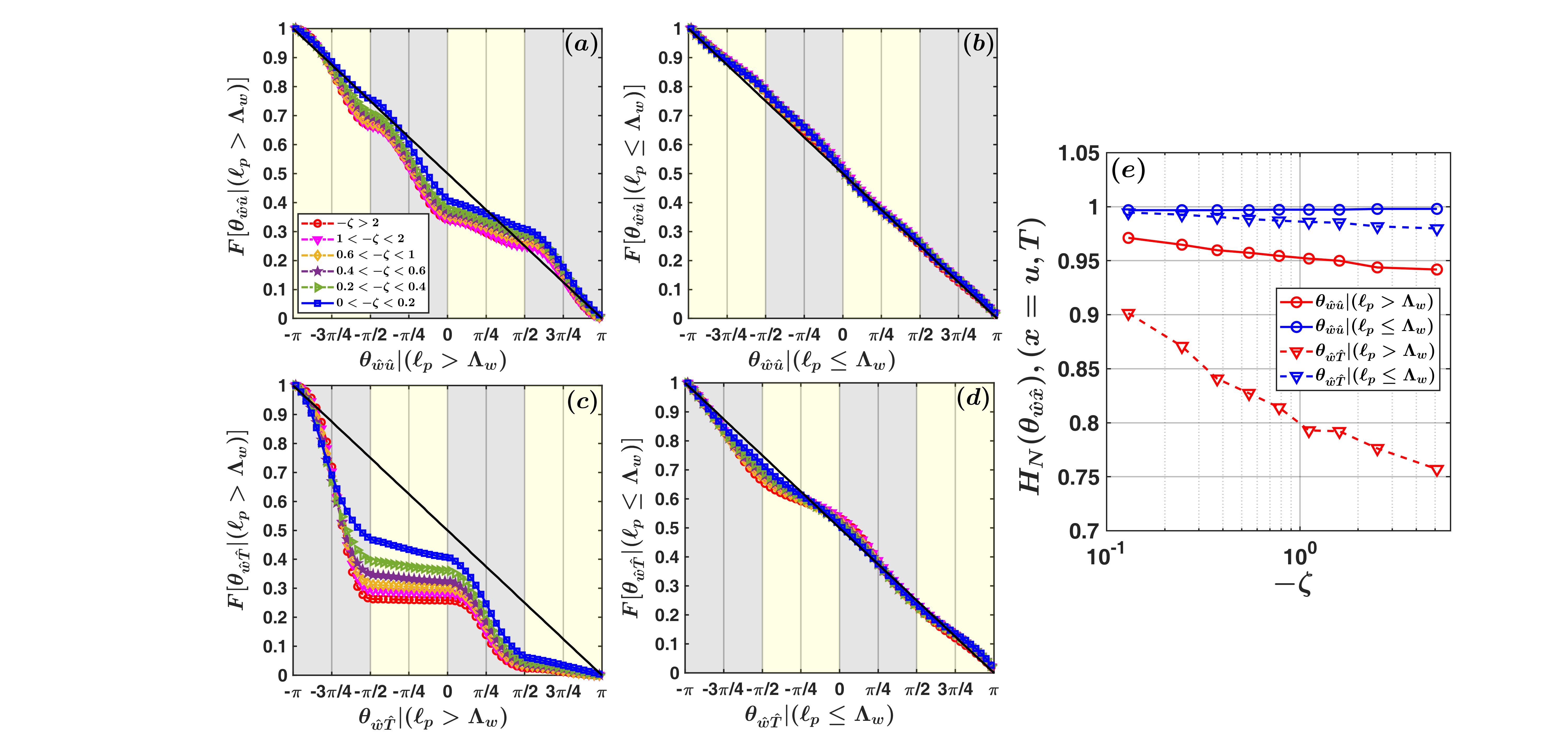}
  \caption{The cumulative distribution functions (CDFs) of the phase angles ($\theta_{\hat{w}\hat{u}}$) for the momentum flux events with length scales (a) $\ell_{p} > \Lambda_{w}$ and (b) $\ell_{p} \leq \Lambda_{w}$. Similarly, the CDFs of the phase angles ($\theta_{\hat{w}\hat{T}}$) are shown for the heat flux events corresponding to (c) $\ell_{p} > \Lambda_{w}$ and (d) $\ell_{p} \leq \Lambda_{w}$. The thick black lines on all the panels denote the CDFs of the phase angles if they were distributed uniformly over the range $-\pi$ to $\pi$. The description of the various markers indicating the six different stability classes is provided in the legend of panel (a). In panel (e), the normalized Shannon entropy associated with the distribution of the phase angles are shown, corresponding to the length scales $\ell_{p} > \Lambda_{w}$ and $\ell_{p} \leq \Lambda_{w}$. The legend of panel (e) describes the markers and the lines on the plot.}
\label{fig:6}
\end{figure}

In order to determine how closely the distribution of the phase angles in figure \ref{fig:5} resemble a uniform distribution, it is imperative to consider the cumulative distribution functions (CDFs), instead of the PDFs. This is because, the CDFs of a uniform distribution can be represented through a linear function. The CDFs of the phase angles and the equivalent CDFs corresponding to a uniform distribution can be written as,
\begin{align}
F(\theta_{\hat{w}\hat{x}})=\int_{\pi}^{\theta_{\hat{w}\hat{x}}} P(\theta_{\hat{w}\hat{x}})\ d\theta_{\hat{w}\hat{x}}
\\
F_{u}(\theta_{\hat{w}\hat{x}})=\frac{\pi-\theta_{\hat{w}\hat{x}}}{2 \pi}
\label{CDF}
\end{align}
where $F(\theta_{\hat{w}\hat{x}})$ is the empirical CDF, and $F_{u}(\theta_{\hat{w}\hat{x}})$ is the CDF for the uniform distribution. Figure \ref{fig:6} shows the CDFs of the phase angles for the large and small scale heat and momentum flux events and compare those with the CDF of a uniform distribution. An apparent observation from figures \ref{fig:6}a and b is, irrespective of stability the statistical characteristics of the CDFs do not seem to differ much from a uniform distribution for both the large and small scale momentum flux events. However, for the large scale heat flux events the CDFs differ significantly from the uniform distribution (figure \ref{fig:6}c), although such contrast becomes less obvious for the small scale heat flux events (figure \ref{fig:6}d). 

Notwithstanding the fact that the CDFs provide a perceptible measure to inspect whether the distributions of the phase angles differ from a uniform distribution, the quantification of such effect remains an issue. To provide a convenient solution to that, we introduce the normalized Shannon entropy of the phase angle distribution corresponding to the large and small scale flux events. From the information theory \citep{banerjee2018coherent}, the normalized Shannon entropy ($H_{N}$) of the phase angle distribution can be defined as,
\begin{equation}
H_{N}(\theta_{\hat{w}\hat{x}})=-\frac{1}{\ln{(N_{\rm b})}}\sum_{i=1}^{N_{\rm b}} P_{i}(\theta_{\hat{w}\hat{x}}) \ln{[P_{i}(\theta_{\hat{w}\hat{x}})]},
\label{shannon}
\end{equation}
where $N_{\rm b}$ is the number of bins in which the $\theta_{\hat{w}\hat{x}}$ values are divided (60 in our case), and $P_{i}(\theta_{\hat{w}\hat{x}})$ is the probability of occurrence of a particular binned value $\theta_{\hat{w}\hat{x}}$. Note that, for a uniform distribution Eq. (\ref{shannon}) will be equal to 1, given $P_{i}(\theta_{\hat{w}\hat{x}})=1/N_{\rm b}$ for all the bin indexes. As a consequence, the departure from unity in Eq. (\ref{shannon}) is regarded as a metric quantifying the discrepancy with a configuration where the phase vectors are randomly oriented. From Eq. (\ref{flux_value_2}) we know that, in the absence of amplitude dependency such configuration of the phase vectors does not cause any transport of the turbulent fluxes. Therefore, within that constraint, the quantity $H_{N}(\theta_{\hat{w}\hat{x}})$ can be considered as a proxy for the transport efficiency. 

Figure \ref{fig:6}e shows the variation of $H_{N}(\theta_{\hat{w}\hat{x}})$, related to both the small and large scale momentum and heat flux events, with the stability ratio $-\zeta$. Before we discuss the relevant features of figure \ref{fig:6}e, it should be noted that the entropies are not computed for single 30-min runs, but for an ensemble of runs within a particular stability bin. This is necessary to ensure that the PDFs of the phase angles are statistically reliable and the estimations of the normalized Shannon entropies are robust. Moreover, in order to document the variation over a substantial range of $-\zeta$ while maintaining the statistical robustness of the results, we divide the $-\zeta$ values into nine number of bins, where each bin contains equal number of 30-min runs (which is 29 for our case, given a total of 261 runs). 

From figure \ref{fig:6}e one could notice that, for all the stability values, the quantity $H_{N}$ of the phase angles remains almost equal to unity ($H_{N} \approx 0.98$), corresponding to both the small scale momentum and heat flux events. Nonetheless, a considerable departure from unity ($H_{N} \approx 0.75$) is noted for the large scale heat flux events, in highly-convective conditions ($-\zeta>1$). Furthermore, there is an apparent tendency that the $H_{N}$ values approach unity ($H_{N} \approx 0.9$) for the large scale heat flux events as the $-\zeta$ values decrease towards the near-neutral stability. Regardless, for the large scale momentum flux events no significant departure from unity is noted in its $H_{N}$ estimates ($H_{N} \approx 0.96$), irrespective of stability.

To put the above described results into perspective, it is useful to discuss them from the standpoint of turbulent structures and the relation with the heat and momentum transport characteristics in a convective ASL flow. From figure \ref{fig:6}, it is evident that for the small scale heat and momentum flux events ($\ell_{p} \leq \Lambda_{w}$), the phase vectors are oriented in a close-to-random manner with the phase angles being distributed in a quasi-uniform way. From the persistence PDFs of the heat and momentum flux events, we know that at scales $\ell_{p} \leq \Lambda_{w}$, they follow a power-law distribution, related to the eddies from the inertial subrange (figures \ref{fig:2} and \ref{fig:3}). According to the Kolmogorov's hypothesis, the turbulence associated with the eddies from the inertial subrange of the spectrum is quasi-isotropic in nature and hardly transport any heat or momentum \citep{tennekes1972first,Wyn72}. Since the random orientation of the phase vectors denotes no flux transport (see Eq. (\ref{flux_value_2})), this complements the properties of turbulence in the inertial subrange, corresponding to scales $\ell_{p} \leq \Lambda_{w}$. 

On the other hand, for the large scale heat flux events ($\ell_{p}>\Lambda_{w}$), associated with the energy containing motions, the distribution of the phase angles differ significantly from the uniform distribution in highly-convective stability. This indicates the vertical velocity and temperature fluctuations are phase-locked to a certain degree in a highly-convective ASL flow and hence related to efficient transport of heat (see Eqs. \ref{flux_value_2} and \ref{flux_value_4}). Such a contention is in agreement with the large-eddy simulation studies, where the researchers have shown that both the vertical velocity and temperature patterns overlay on each other in the form of cellular structures \citep{khanna1998three,salesky2017nature}. This configuration is efficient in transporting the heat flux, which agrees with our assessment. Nevertheless, the phase angle PDFs between the $w^{\prime}$ and $T^{\prime}$ signals gradually resemble a uniform distribution as the near-neutral stability is approached. This behaviour is in concurrence with the observation that the heat-transport efficiency decreases as the turbulence becomes more shear dominated \citep{li2011coherent}. Therefore, we can conclude that the heat-transport efficiency in a convective ASL flow can be explained by the departure of the phase-angles from a uniform distribution, corresponding to the events with scales $\ell_{p}>\Lambda_{w}$. 

Interestingly, an identical deduction cannot be made for the large scale momentum flux events ($\ell_{p}>\Lambda_{w}$). This is because, for these events, the phase angle PDFs remain similar to a uniform distribution for all the $-\zeta$ values. From Eq. \ref{flux_value_4} it would imply that the momentum transport remains inefficient irrespective of the strength of the thermal stratification. However, such an implication does not agree with the ubiquitous result that the momentum transport efficiency increases with the decrease in $-\zeta$. Needless to say, this brings into consideration the role of amplitudes in the momentum flux generation. While connecting the phase angle PDFs directly with the correlation coefficient $R_{uw}$, it is assumed that the amplitudes of all the phase vectors can be replaced with a single value, while preserving their angles. Since such an assumption produces output which is incompatible with the measurements, we thus infer that the amplitude variations play a significant part to generate the momentum flux. Ignoring the amplitude effect results in almost no transport of averaged momentum even during the near-neutral conditions. This agrees with the observations of \citet{haugen1971experimental}, \citet{hogstrom1996organized}, and \citet{narasimha2007turbulent} where they noted that in a near-neutral ASL flow, the averaged momentum flux is mainly generated through burst like events associated with strong gusts in the streamwise velocity fluctuations. We present our conclusions in Section \ref{conclusion}. 

\section{Conclusion}
\label{conclusion}
In this study, we provide a detailed account of the persistence properties of turbulent heat and momentum fluxes as obtained from the SLTEST experimental dataset, in a convective surface layer. Furthermore, we also establish a novel linkage between the persistence of the flux events and the heat and momentum transport characteristics. We develop such correspondence through a framework based on the concept of phase space in non-linear dynamical systems. 

On a larger scale, the ramifications of this research are directed towards providing answers to the questions posed in the introduction. Keeping that in mind, the important results from this paper can be noted as: 
\begin{enumerate}
    \item The comparison of the persistence PDFs of heat and momentum fluxes ($u^{\prime}w^{\prime}$ and $w^{\prime}T^{\prime}$) with the component signals ($u^{\prime}$ and $w^{\prime}$ or $w^{\prime}$ and $T^{\prime}$) reveals that at scales ($\ell_{p}$) smaller than the integral scale of the vertical velocity ($\Lambda_{w}$), the persistence PDFs of the component signals and their products are in excellent agreement with each other. For such scales, the persistence PDFs of the products ($u^{\prime}w^{\prime}$ and $w^{\prime}T^{\prime}$) follow an identical power-law distribution with an exponent of $-1.4$, irrespective of the stability conditions. On the other hand, for scales larger than the integral scale of the vertical velocity, both the flux persistence PDFs nearly collapse on the persistence PDFs of the $w^{\prime}$ signals and deviate significantly from the $u^{\prime}$ or $T^{\prime}$ signals. 
    \item The flux persistence PDFs are investigated separately by considering the distribution of the time scales from the four different quadrants. We discover that, for the momentum flux events, the persistence PDFs of the $u^{\prime}w^{\prime}$ signals are indistinguishable for all the four quadrants by collapsing onto one another. However, for the heat flux events, the effect of quadrant separation on the $w^{\prime}T^{\prime}$ persistence PDFs remains insignificant for scales smaller than the integral scale of the vertical velocity. Contrarily, for $\ell_{p}>\Lambda_{w}$, the persistence PDFs of the heat flux events are primarily governed by the down-gradient quadrants; warm-updrafts and cold-downdrafts.
    \item For scales $\ell_{p} \leq \Lambda_{w}$, the persistence PDFs of both the flux events show a similar characteristics in terms of quadrant behaviour and are also akin to a power-law distribution. Such power-law behaviour is related to the eddies from the inertial subrange of the turbulence spectrum. Nevertheless, at scales $\ell_{p}>\Lambda_{w}$, the persistence PDFs differ from the power-law distribution and drop off exponentially. Moreover, at such time scales, the persistence PDFs of the $w^{\prime}T^{\prime}$ signals are dominated by the organized heat flux events from the down-gradient quadrants. Therefore, the investigation of flux persistence leads to a criterion to separate two different eddy processes, based on the integral scale of $w^{\prime}$. Since the integral length scale of $w^{\prime}$ is of the same order as $z$, they reflect the properties of the attached and detached eddies, based on the Townsend's attached eddy model. 
    \item To gain insight into the transport mechanisms related to these two eddy processes, we scrutinize the effects of phase angles and amplitudes associated with the flux component signals (\{$u^{\prime}$,$w^{\prime}$\} and \{$T^{\prime}$,$w^{\prime}$\}). We derive a simple relation between the PDFs of the phase angles and transport efficiencies by ignoring the variations in the amplitudes. Under such assumption, the departure of the phase angle PDFs from a uniform distribution is shown to be a necessary condition for the flux transport. The results indicate that the phase angles of the component signals related to the heat and momentum flux events with length scales $\ell_{p} \leq \Lambda_{w}$ are almost uniformly distributed. Given these events are associated with the detached eddies from the inertial subrange, this agrees with the general notion that quasi-isotropic turbulence for such size ranges hardly transport any flux.
    \item For the large scale heat flux events ($\ell_{p}>\Lambda_{w}$), the departure of the phase angle PDFs from a uniform distribution is the strongest for the highly-convective regime. However, with the change in stability towards the near-neutral regime, the same PDFs resemble closely a uniform distribution. Such variation explains the gradual reduction in the heat transport efficiency as the near-neutral stability is approached.  
    \item On the other hand, for the large scale momentum flux events, the phase angle PDFs remain close to a uniform distribution irrespective of all the stability classes. Such a behaviour would suggest, on an average there is nearly no transport of momentum in a convective ASL flow. This result is antithetical to the observation that the momentum transport efficiency increases with the decrease in strength of the thermal stratification. In order to explain the contradiction, the amplitude effects need to be taken into account for the momentum transport.  
    \end{enumerate}

To sum up, we analyze the heat and momentum transport characteristics of two different eddy processes, identified through the statistical properties of the flux persistence PDFs in a convective ASL flow. We demonstrate that the heat transport efficiency in convective conditions is related to the phase angle distributions, associated with the events which persist for times larger than the integral scale of vertical velocity. On the contrary, for the momentum transport efficiency, the phase information between the streamwise and vertical velocity fluctuations remains largely irrelevant. More importantly, if the amplitude effects are disregarded then it results in a situation with nearly no transport of streamwise momentum. For practical purposes, these results offer a unique perspective towards modelling the heat and momentum transport processes in a convective ASL flow.

\section*{Acknowledgements}
Indian Institute of Tropical Meteorology (IITM) is an autonomous institute fully funded by the Ministry of Earth Sciences, Government of India. The authors S Chowdhuri and T.V.Prabha gratefully acknowledge the Director, IITM, Dr. Kalm\'{a}r-Nagy, and Mr. Siddharth Kumar for the constant encouragement and stimulating discussions during the research. The author T. Banerjee acknowledges the funding support from the University of California Laboratory Fees Research Program funded by the UC Office of the President (UCOP), grant ID LFR-20-653572. Additional support was provided by the new faculty start up grant provided by the Department of Civil and Environmental Engineering, and the Henry Samueli School of Engineering, University of California, Irvine. The authors are also immensely grateful to Dr. Keith G McNaughton for kindly providing them the SLTEST dataset for this research. The computer codes used in this study are available to all the researchers by contacting the corresponding author.  

\bibliographystyle{apalike}  
\bibliography{references}

\begin{thebibliography}{}

\bibitem[Alfredsson and Johansson, 1984]{alfredsson1984time}
Alfredsson, P. and Johansson, A. (1984).
\newblock Time scales in turbulent channel flow.
\newblock {\em Phys. Fluids}, 27(8):1974--1981.

\bibitem[Baldocchi et~al., 2001]{baldocchi2001fluxnet}
Baldocchi, D., Falge, E., Gu, L., Olson, R., Hollinger, D., Running, S.,
  Anthoni, P., Bernhofer, C., Davis, K., Evans, R., et~al. (2001).
\newblock Fluxnet: A new tool to study the temporal and spatial variability of
  ecosystem-scale carbon dioxide, water vapor, and energy flux densities.
\newblock {\em Bull. Am. Meteorol. Soc.}, 82(11):2415--2434.

\bibitem[Banerjee et~al., 2018]{banerjee2018coherent}
Banerjee, T., Vercauteren, N., Muste, M., and Yang, D. (2018).
\newblock Coherent structures in wind shear induced
  wave--turbulence--vegetation interaction in water bodies.
\newblock {\em Agr. Forest Meteorol.}, 255:57--67.

\bibitem[Bershadskii et~al., 2004]{bershadskii2004clusterization}
Bershadskii, A., Niemela, J., Praskovsky, A., and Sreenivasan, K. (2004).
\newblock “{Clusterization}” and intermittency of temperature fluctuations
  in turbulent convection.
\newblock {\em Phys. Rev. E}, 69(5):056314.

\bibitem[Caramori et~al., 1994]{caramori1994structural}
Caramori, P., Schuepp, P., Desjardins, R., and MacPherson, I. (1994).
\newblock Structural analysis of airborne flux estimates over a region.
\newblock {\em J. Clim.}, 7(5):627--640.

\bibitem[Cava et~al., 2012]{cava2012role}
Cava, D., Katul, G., Molini, A., and Elefante, C. (2012).
\newblock The role of surface characteristics on intermittency and
  zero-crossing properties of atmospheric turbulence.
\newblock {\em J. Geophys. Res. Atmos.}, 117(D1).

\bibitem[Chowdhuri et~al., 2020a]{chowdhuri2020persistence}
Chowdhuri, S., Kalmár-Nagy, T., and Banerjee, T. (2020a).
\newblock Persistence analysis of velocity and temperature fluctuations in
  convective surface layer turbulence.
\newblock {\em Phys. Fluids}, 32(7):076601.

\bibitem[Chowdhuri et~al., 2020b]{chowdhuri2019revisiting}
Chowdhuri, S., Kumar, S., and Banerjee, T. (2020b).
\newblock Revisiting the role of intermittent heat transport towards {Reynolds}
  stress anisotropy in convective turbulence.
\newblock {\em J. Fluid Mech.}, 899:A26.

\bibitem[Chowdhuri and Prabha, 2019]{chowdhuri2019evaluation}
Chowdhuri, S. and Prabha, T. (2019).
\newblock An evaluation of the dissimilarity in heat and momentum transport
  through quadrant analysis for an unstable atmospheric surface layer flow.
\newblock {\em Environ. Fluid Mech.}, 19(2):513--542.

\bibitem[Davidson, 2015]{davidson2015turbulence}
Davidson, P.~A. (2015).
\newblock {\em Turbulence: an introduction for scientists and engineers}.
\newblock Oxford university press.

\bibitem[De~Bruin et~al., 1993]{de1993verification}
De~Bruin, H., Kohsiek, W., and Van Den~Hurk, B. (1993).
\newblock A verification of some methods to determine the fluxes of momentum,
  sensible heat, and water vapour using standard deviation and structure
  parameter of scalar meteorological quantities.
\newblock {\em Boundary-Layer Meteorol.}, 63(3):231--257.

\bibitem[Duncan and Schuepp, 1992]{duncan1992method}
Duncan, M. and Schuepp, P. (1992).
\newblock A method to delineate extreme structures within airborne flux traces
  over the {FIFE} site.
\newblock {\em J. Geophys. Res. Atmos.}, 97(D17):18487--18498.

\bibitem[Haugen et~al., 1971]{haugen1971experimental}
Haugen, D., Kaimal, J., and Bradley, E. (1971).
\newblock An experimental study of {Reynolds} stress and heat flux in the
  atmospheric surface layer.
\newblock {\em Q. J. R. Meteorol. Soc.}, 97(412):168--180.

\bibitem[H{\"o}gstr{\"o}m and Bergstr{\"o}m, 1996]{hogstrom1996organized}
H{\"o}gstr{\"o}m, U. and Bergstr{\"o}m, H. (1996).
\newblock Organized turbulence structures in the near-neutral atmospheric
  surface layer.
\newblock {\em J. Atmos. Sci.}, 53(17):2452--2464.

\bibitem[Holmes et~al., 2012]{holmes2012turbulence}
Holmes, P., Lumley, J.~L., Berkooz, G., and Rowley, C.~W. (2012).
\newblock {\em Turbulence, coherent structures, dynamical systems and
  symmetry}.
\newblock Cambridge university press.

\bibitem[Kailasnath and Sreenivasan, 1993]{kailasnath1993zero}
Kailasnath, P. and Sreenivasan, K. (1993).
\newblock Zero crossings of velocity fluctuations in turbulent boundary layers.
\newblock {\em Phys. Fluids}, 5(11):2879--2885.

\bibitem[Katul et~al., 1994]{katul1994conditional}
Katul, G., Albertson, J., Parlange, M., Chu, C., and Stricker, H. (1994).
\newblock Conditional sampling, bursting, and the intermittent structure of
  sensible heat flux.
\newblock {\em J. Geophys. Res. Atmos.}, 99(D11):22869--22876.

\bibitem[Khanna and Brasseur, 1998]{khanna1998three}
Khanna, S. and Brasseur, J. (1998).
\newblock Three-dimensional buoyancy-and shear-induced local structure of the
  atmospheric boundary layer.
\newblock {\em J. Atmos. Sci.}, 55(5):710--743.

\bibitem[Kline et~al., 1967]{kline1967structure}
Kline, S.~J., Reynolds, W.~C., Schraub, F., and Runstadler, P. (1967).
\newblock The structure of turbulent boundary layers.
\newblock {\em J. Fluid Mech.}, 30(4):741--773.

\bibitem[Li and Bou-Zeid, 2011]{li2011coherent}
Li, D. and Bou-Zeid, E. (2011).
\newblock Coherent structures and the dissimilarity of turbulent transport of
  momentum and scalars in the unstable atmospheric surface layer.
\newblock {\em Boundary-Layer Meteorol.}, 140(2):243--262.

\bibitem[Lu and Willmarth, 1973]{lu1973measurements}
Lu, S. and Willmarth, W. (1973).
\newblock Measurements of the structure of the {Reynolds} stress in a turbulent
  boundary layer.
\newblock {\em J. Fluid Mech.}, 60(3):481--511.

\bibitem[Luchik and Tiederman, 1987]{luchik1987timescale}
Luchik, T. and Tiederman, W. (1987).
\newblock Timescale and structure of ejections and bursts in turbulent channel
  flows.
\newblock {\em J. Fluid Mech.}, 174:529--552.

\bibitem[Mahrt and Paumier, 1984]{mahrt1984heat}
Mahrt, L. and Paumier, J. (1984).
\newblock Heat transport in the atmospheric boundary layer.
\newblock {\em J. Atmos. Sci.}, 41(21):3061--3075.

\bibitem[Majumdar, 1999]{majumdar1999persistence}
Majumdar, S.~N. (1999).
\newblock Persistence in nonequilibrium systems.
\newblock {\em Curr. Sci.}, pages 370--375.

\bibitem[Marusic and Monty, 2019]{marusic2019attached}
Marusic, I. and Monty, J.~P. (2019).
\newblock Attached eddy model of wall turbulence.
\newblock {\em Annu. Rev. Fluid Mech.}, 51:49--74.

\bibitem[Monin and Yaglom, 1971]{monin2013statistical}
Monin, A. and Yaglom, A. (1971).
\newblock {\em Statistical fluid mechanics: mechanics of turbulence}, volume~1.
\newblock MIT Press.

\bibitem[Narasimha, 1990]{narasimha1990utility}
Narasimha, R. (1990).
\newblock The utility and drawbacks of traditional approaches.
\newblock In {\em Whither Turbulence? Turbulence at the Crossroads}, pages
  13--48. Springer.

\bibitem[Narasimha and Kailas, 1990]{narasimha1990turbulent}
Narasimha, R. and Kailas, S. (1990).
\newblock Turbulent bursts in the atmosphere.
\newblock {\em Atmos. Environ.}, 24(7):1635--1645.

\bibitem[Narasimha et~al., 2007]{narasimha2007turbulent}
Narasimha, R., Kumar, S., Prabhu, A., and Kailas, S. (2007).
\newblock Turbulent flux events in a nearly neutral atmospheric boundary layer.
\newblock {\em Phil. Trans. R. Soc. A}, 365(1852):841--858.

\bibitem[Rao et~al., 1971]{rao1971bursting}
Rao, K., Narasimha, R., and Badri~Narayanan, M. (1971).
\newblock The `bursting' phenomenon in a turbulent boundary layer.
\newblock {\em J. Fluid Mech.}, 48(2):339--352.

\bibitem[Salesky et~al., 2017]{salesky2017nature}
Salesky, S., Chamecki, M., and Bou-Zeid, E. (2017).
\newblock On the nature of the transition between roll and cellular
  organization in the convective boundary layer.
\newblock {\em Boundary-Layer Meteorol.}, 163(1):41--68.

\bibitem[Shang et~al., 2003]{shang2003measured}
Shang, X., Qiu, X., Tong, P., and Xia, K. (2003).
\newblock Measured local heat transport in turbulent {Rayleigh}-{B{\'e}nard}
  convection.
\newblock {\em Phys. Rev. Lett.}, 90(7):074501.

\bibitem[Sreenivasan et~al., 1983]{sreenivasan1983zero}
Sreenivasan, K., Prabhu, A., and Narasimha, R. (1983).
\newblock Zero-crossings in turbulent signals.
\newblock {\em J. Fluid Mech.}, 137:251--272.

\bibitem[Taylor, 1938]{Tay38}
Taylor, G.~I. (1938).
\newblock The spectrum of turbulence.
\newblock {\em Proc. Roy. Soc. London A}, 164(919):476--490.

\bibitem[Tennekes and Lumley, 1972]{tennekes1972first}
Tennekes, H. and Lumley, J. (1972).
\newblock {\em A first course in turbulence}.
\newblock MIT press.

\bibitem[Townsend, 1976]{townsend1980structure}
Townsend, A. (1976).
\newblock {\em The structure of turbulent shear flow}.
\newblock Cambridge university press.

\bibitem[Wallace, 2016]{wallace2016quadrant}
Wallace, J. (2016).
\newblock Quadrant analysis in turbulence research: history and evolution.
\newblock {\em Annu. Rev. Fluid Mech.}, 48:131--158.

\bibitem[Wallace et~al., 1972]{wallace1972wall}
Wallace, J.~M., Eckelmann, H., and Brodkey, R.~S. (1972).
\newblock The wall region in turbulent shear flow.
\newblock {\em J. Fluid Mech.}, 54(1):39--48.

\bibitem[Wyngaard and Cot{\'e}, 1972]{Wyn72}
Wyngaard, J. and Cot{\'e}, O. (1972).
\newblock Cospectral similarity in the atmospheric surface layer.
\newblock {\em Q. J. Roy. Meteorol. Soc.}, 98(417):590--603.

\bibitem[Wyngaard, 1992]{wyngaard1992atmospheric}
Wyngaard, J.~C. (1992).
\newblock Atmospheric turbulence.
\newblock {\em Annu. Rev. Fluid Mech.}, 24(1):205--234.

\end{thebibliography}
\clearpage

\section*{Supplementary material}
\renewcommand{\thefigure}{S\arabic{figure}}
\setcounter{figure}{0}

\begin{figure}[h]
\centering
\hspace*{-0.75in}
\includegraphics[width=1.25\textwidth]{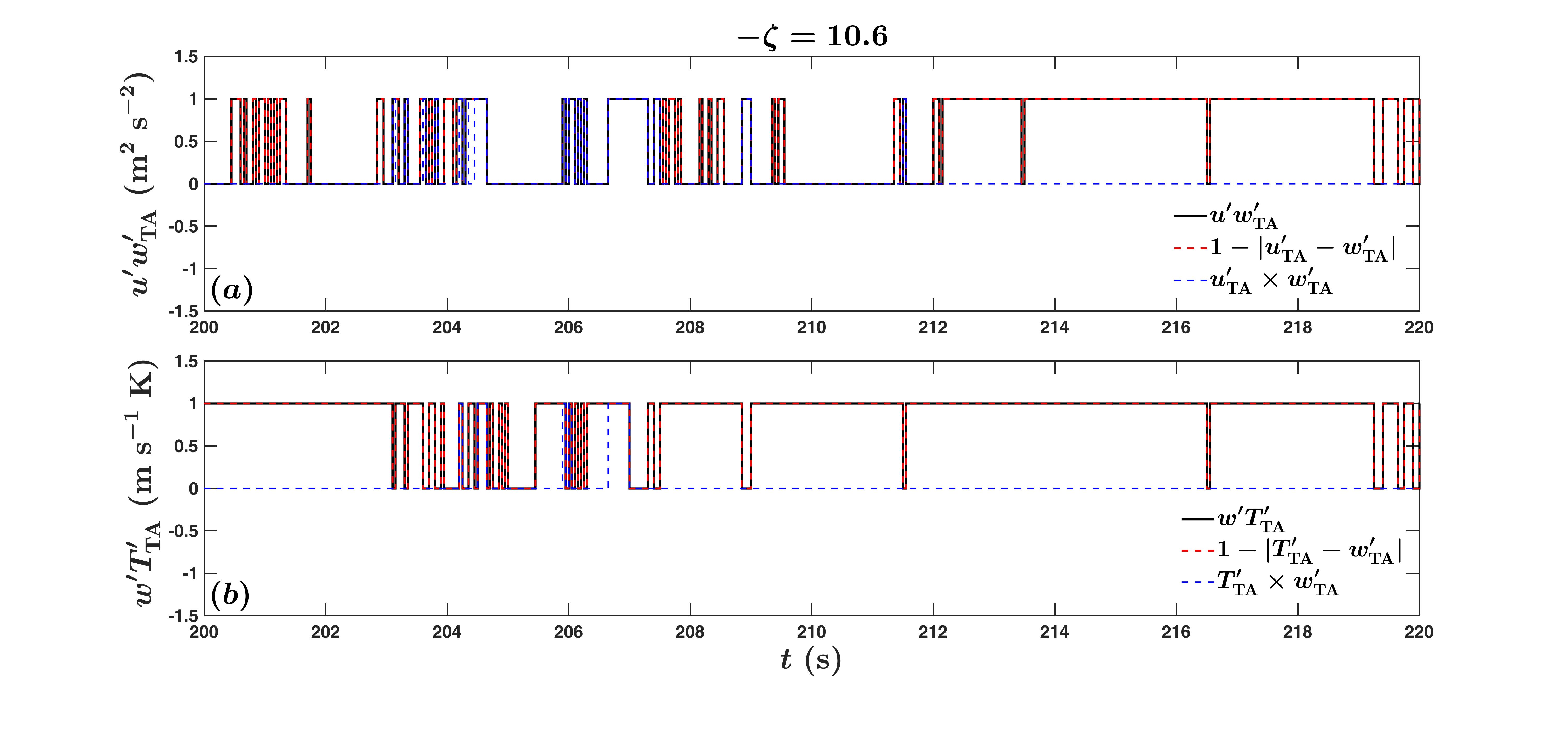}
 \caption{A typical example is shown for the original TA approximated time series of (a) $u^{\prime}w^{\prime}$ and (b) $w^{\prime}T^{\prime}$ for a highly-convective stability, corresponding to $-\zeta=10.6$. In both the panels, the original $(x^{\prime}w^{\prime})_{\rm TA}$ (thick black lines) are compared with the formulations $1-|x^{\prime}_{\rm TA}-w^{\prime}_{\rm TA}|$ (dashed red lines) and $x^{\prime}_{\rm TA} \times w^{\prime}_{\rm TA}$ (dashed blue lines), where $x$ can be either $u$ or $T$. It can be noted that, the expression $1-|x^{\prime}_{\rm TA}-w^{\prime}_{\rm TA}|$ captures all the signature of $(x^{\prime}w^{\prime})_{\rm TA}$ perfectly well.}
\label{fig:s1}
\end{figure}

\begin{figure}[h]
\centering
\hspace*{-1.5in}
\includegraphics[width=1.3\textwidth]{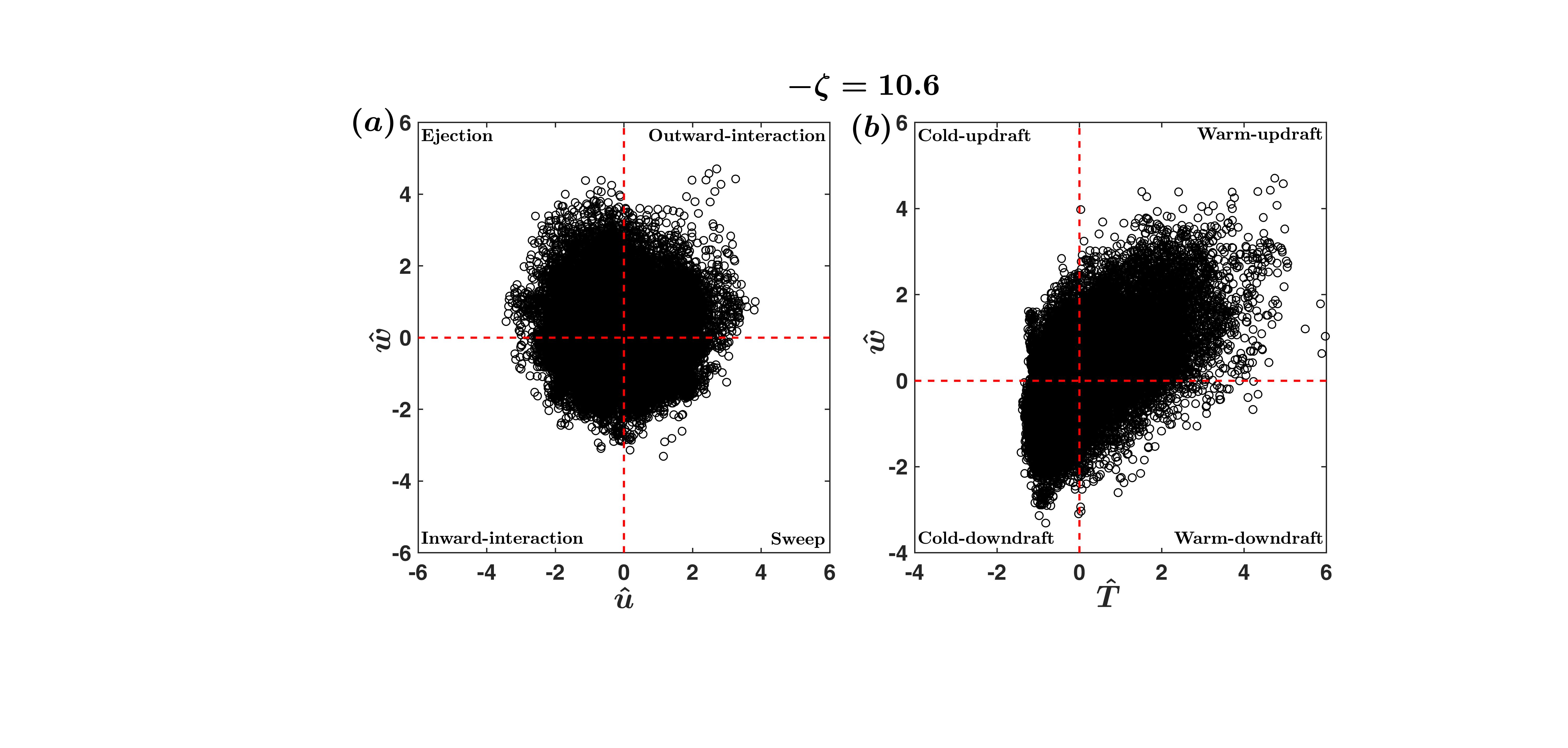}
\vspace{-2cm}
 \caption{An example is shown to illustrate the concepts of (a) $\hat{u}$-$\hat{w}$ and (b) $\hat{T}$-$\hat{w}$ quadrant planes. Note that, the symbol $\hat{x}=x^{\prime}/\sigma_{x}$, where $x$ can be $u$, $w$, or $T$. The points shown in black belong to a 30-min run from highly-convective stability, corresponding to $-\zeta=10.6$. The four different quadrants for each of the planes are marked in the respective panels.}
\label{fig:s2}
\end{figure}

\end{document}